\documentclass[a4,12pt]{article}%
\usepackage{graphicx}
\usepackage{amsmath}
\usepackage{amsfonts}
\usepackage{amssymb}
\usepackage{epic}%
\newtheorem{theorem}{Theorem}[section]

\newtheorem{corollary}[theorem]{Corollary}

\newtheorem{definition}[theorem]{Definition}

\newtheorem{lemma}[theorem]{Lemma}

\newtheorem{proposition}[theorem]{Proposition}

\newenvironment{proof}[1][Proof]{\textbf{#1.} }{\ \rule{0.5em}{0.5em}}
\renewcommand{\theequation}{\thesection.\arabic{equation}}
\begin{document}

\title{Surface transitions of the semi-infinite Potts model I: the high bulk
temperature regime}
\author{C. Dobrovolny$^{1}$, L. Laanait$^{2}$, and J. Ruiz$^{3}$ }
\date{}
\maketitle

\renewcommand{\theequation}{\thesection.\arabic{equation}}
\renewcommand{\thefootnote}{} \footnote{Preprint CPT--2002/P.4465 revised,
published in J.\ Stat.\ Phys.\ \textbf{114} 1269--1302 (2004)}
\renewcommand{\thefootnote}{\arabic{footnote}} \footnotetext[1]{CPT, CNRS,
Luminy case 907, F-13288 Marseille Cedex 9, France. \newline E-mail:
\textit{dodrovol@cpt.univ-mrs.fr}} \footnotetext[2]{Ecole Normale
sup\'{e}rieure de Rabat, B.P. 5118 Rabat, Morocco \newline E-mail:
\textit{laanait@yahoo.fr}} \footnotetext[3]{CPT, CNRS, Luminy case 907,
F-13288 Marseille Cedex 9, France.\newline E-mail:
\textit{ruiz@cpt.univ-mrs.fr}} \setcounter{footnote}{3} \thispagestyle{empty}

\begin{quote}
{\footnotesize \textsc{Abstract:} We propose a rigorous approach of
Semi-Infinite lattice systems illustrated with the study of surface
transitions of the semi-infinite Potts model. }

\vskip15pt

{\footnotesize \textsc{Key words:} Phase transitions, Potts model,
Semi-infinite lattice systems.}
\end{quote}


\newpage

\section{Introduction and definitions}

\setcounter{equation}{0}

\subsection{Introduction}

Semi-infinite lattice systems with free surface were studied extensively by
mean field and numerical approaches. They exhibit a rich variety of critical
behaviour depending on the relative strengths of the surface and bulk coupling
constants. \textit{Ordinary critical behaviour} occurs when the surface spins
order at the bulk critical temperature as the temperature is lowered. When the
surface couplings are sufficiently enhanced, the surface spins order above the
bulk critical temperature in a transition known as the \textit{surface
transition}. As the temperature is then lowered through the bulk critical
temperature the \textit{extraordinary transition} takes place. Finally, a
\textit{special} critical behaviour arises if the couplings are adjust such
that the bulk and surface transitions coincide. We refer the reader to Binder
\cite{B} for a review on the subject. Exact results may be found in \cite{A},
\cite{MW}. A rigorous approach of the semi-infinite Ising model is proposed in
a series of articles by Fr\"{o}hlich and Pfister, \cite{FP}-\cite{FP3}. In
particular, they observed that the surface transition does not occur for
temperatures less than the critical bulk temperature (see Figure 1 in
\cite{FP}).

A new scheme, was found by Lipowsky \cite{Li}, in the case of the $q$-state
Potts model. In the many component limit $q\rightarrow\infty$, both mean field
theory and Migdal-Kadanoff renormalization group predict a new phase where the
bulk is ordered while the free surface is disordered. The relative strengths
of the surface and bulk couplings can be chosen such that when the temperature
is lowered, first the bulk spins order through an extraordinary transition and
then surface spins order through a surface transition, or a new special
transition arises (see diagram Figure 1 in \cite{Li}). Intuitively, one can
think that the low temperature ordered bulk acts upon the spin on the boundary
surface like an external magnetic field. Roughly speaking, one can imagine
that this effect modifies the Hamiltonian in the boundary surface by a term
containing a magnetic field. Then for the Ising case a surface phase
transition cannot occur when the bulk is ordered. However, this observation is
not true for the q-state Potts model where it is known for $q$ large that a
disordered state arises even in a presence of a magnetic field and coexists
with the ordered one aligned with the magnetic field \cite{BBL}.

The aim of this work is to prove that for $q$ large the mean field or
renormalization group prediction are indeed correct . We will study both
surface phase transitions (of the semi-infinite $q$-state Potts model for bulk
dimension $d\geq3$), using the Fortuin-Kasteleyn representation of this model.
In doing so we think that the method can be extended to a broader class of systems.

Let us sketch out the main scheme of the approach. Since we have to study the
surface phase transitions and thus the possible coexistence of
\textquotedblleft surface phases\textquotedblright\ (that actually holds), we
define following \cite{FP} two surface free energies corresponding
respectively to the high and low temperature bulk. Here by \textquotedblleft
surface phases\textquotedblright\ we mean states invariant under horizontal
translations. The ratio of partition functions defining the surface free
energies is then expanded using high or low convergent expansion. This leads
to horizontally translation invariant systems called hydra models that are
then analyzed with the help of Pirogov-Sinai theory \cite{S}. Actually, the
theory developed for the study of translation invariant bulk states can be
implemented to the case under consideration.

The paper is organized as follow. The surface free energies are defined in the
next subsection. In Section 2, we present the Fortuin-Kasteleyn representation
and the ground state analysis of the model. Then we give convergent cluster
expansions of the bulk partition function in the high and low (bulk)
temperature regime. The hydra system for the high temperature bulk case is
presented in Subsection 3.1. Subsections 3.2 and 3.3 are devoted to the large
q analysis of the hydra model. Finally, some proofs are postponed to the appendix.

\subsection{Surface free energies \label{S:1.1}}

Consider a ferromagnetic Potts model on the semi-infinite lattice
$\mathbb{L}=\mathbb{Z}^{d-1}\times\mathbb{Z}^{+}$ of dimension $d\geq3$. At
each site $i=\left\{  i_{1},...,i_{d}\right\}  \in\mathbb{L}$, with
$i_{\alpha}\in\mathbb{Z}$ for $\alpha=1,...,d-1$ and $i_{d}\in\mathbb{Z}^{+}$,
there is a spin variable $\sigma_{i}$ taking its values in the set
$\mathcal{Q}\equiv\{0,1,\ldots,q-1\}$. We let $d(i,j)=\max_{\alpha
=1,...,d}\left\vert i_{\alpha}-j_{\alpha}\right\vert $ be the distance between
two sites, $d(i,\Omega)=\min_{j\in\Omega}d(i,j)$ be the distance between the
site $i$ and a subset $\Omega\subset\mathbb{L}$, and $d(\Omega,\Omega^{\prime
})=\min_{i\in\Omega,j\in\Omega^{\prime}}d(i,j)$ be the distance between two
subsets of $\mathbb{L}$ . The Hamiltonian of the system is given by
\begin{equation}
H\equiv-\sum_{\langle i,j\rangle}K_{ij}\delta(\sigma_{i},\sigma_{j})
\label{eq:1.1}%
\end{equation}
where the sum runs over nearest neighbor pairs $\langle i,j\rangle$ (i.e. at
Euclidean distance $d_{\text{E}}(i,j)=1$) of a finite subset $\Omega
\subset\mathbb{L}$, and $\delta$ is the Kronecker symbol: $\delta(\sigma
_{i},\sigma_{j})=1$ if $\sigma_{i}=\sigma_{j}$, and $0$ otherwise. The
coupling constants $K_{ij}$ can take two values according both $i$ and $j$
belong to the \emph{boundary layer} $\mathbb{L}_{0}\equiv\{i\in\mathbb{L}\mid
i_{d}=0\}$, or not:
\begin{equation}
K_{ij}=\left\{
\begin{array}
[c]{l}%
K>0\hspace{0.35cm}\text{if}\quad\langle i,j\rangle\subset\mathbb{L}_{0}\\
J>0\hspace{0.35cm}\text{otherwise}%
\end{array}
\right.  \label{eq:1.2}%
\end{equation}

The partition function is defined by:
\begin{equation}
Z^{p}(\Omega)\equiv\sum e^{-\beta H}\chi_{\Omega}^{p} \label{eq:1.3}%
\end{equation}
Here the sum is over configurations $\sigma_{\Omega}\in\mathcal{Q}^{\Omega}$,
$\beta$ is the inverse temperature, and $\chi_{\Omega}^{p}$ is a
characteristic function giving the boundary conditions. In particular, we will
consider the following boundary conditions:

\begin{itemize}
\item the free boundary condition: $\chi_{\Omega}^{\text{f}}=1$

\item the ordered boundary condition: $\chi_{\Omega}^{\text{o}}=\prod
_{i\in\partial\Omega}\delta(\sigma_{i},0)$, where the boundary of $\Omega$ is
the set of sites of $\Omega$ at distance one to its complement $\partial
\Omega=\left\{  i\in\Omega:d(i,\mathbb{L}\setminus\Omega)=1\right\}  $.

\item the free boundary condition in the bulk and ordered boundary condition
on the surface: $\chi_{\Omega}^{\text{fo}}=\prod_{i\in\partial_{s}\Omega
}\delta(\sigma_{i},0)$, where $\partial_{s}\Omega\equiv\partial\Omega
\cap\mathbb{L}_{0}$

\item the ordered boundary condition in the bulk and free boundary condition
on the surface: $\chi_{\Omega}^{\text{of}}=\prod_{i\in\partial_{b}\Omega
}\delta(\sigma_{i},0)$,where $\partial_{b}\Omega=\partial\Omega\setminus
\partial_{s}\Omega$.
\end{itemize}

Let us now consider the finite box
\[
\Omega=\{i\in\mathbb{L}\mid\max_{\alpha=1,...,d-1}|i_{\alpha}|\leq L,\;0\leq
i_{d}\leq M\}
\]
its projection $\Sigma=\Omega\cap\mathbb{L}_{0}=\{i\in\Omega\mid i_{d}=0\}$ on
the boundary layer and its bulk part $\Lambda=\Omega\backslash\Sigma
=\{i\in\Omega\mid1\leq i_{d}\leq M\}$.

Following Fr\"{o}hlich and Pfister \cite{FP} we introduce two surface free
energies, the \emph{free (or disordered) surface free energy\/} with free
boundary condition in the bulk,
\begin{equation}
g_{\text{f}}=-\lim_{L\rightarrow\infty}\frac{1}{|\Sigma|}\lim_{M\rightarrow
\infty}\ln\frac{Z^{\text{f}}(\Omega)}{Q^{\text{f}}(\Lambda)} \label{eq:1.4}%
\end{equation}
and the \emph{ordered surface free energy\/} corresponding to ordered boundary
condition in the bulk,
\begin{equation}
g_{\text{o}}=-\lim_{L\rightarrow\infty}\frac{1}{|\Sigma|}\lim_{M\rightarrow
\infty}\ln\frac{Z^{\text{o}}(\Omega)}{Q^{\text{o}}(\Lambda)} \label{eq:1.5}%
\end{equation}
Here $|\Sigma|=(2L+1)^{d-1}$ is the number of lattice site in $\Sigma$, and
$Q^{\text{f}}(\Lambda)$ and $Q^{\text{o}}(\Lambda)$ are the following bulk
partition functions: \arraycolsep2pt
\begin{align}
Q^{\text{f}}(\Lambda)  &  =\sum\exp\Big\{\beta J\sum_{\langle i,j\rangle
\subset\Lambda}\delta(\sigma_{i},\sigma_{j})\Big\}\label{eq:1.6}\\
Q^{\text{o}}(\Lambda)  &  =\sum\exp\Big\{\beta J\sum_{\langle i,j\rangle
\subset\Lambda}\delta(\sigma_{i},\sigma_{j})\Big\}\prod_{i\in\partial\Lambda
}\delta(\sigma_{i},0) \label{eq:1.7}%
\end{align}
where the two sums are over configurations $\sigma_{\Lambda}\in\mathcal{Q}%
^{\Lambda}$. The surface free energies do not depend on the boundary condition
on the surface, in particular one can replace $Z^{\text{f}}(\Omega)$ by
$Z^{\text{fo}}(\Omega)$ in (\ref{eq:1.4}) and replace $Z^{\text{o}}(\Omega)$
by $Z^{\text{of}}(\Omega)$ in (\ref{eq:1.5}). The partial derivative of the
surface free energy with respect to $\beta K$ represents the mean surface
energy. As a result of this paper we get, for $q$ large and 
$e^{\beta J}-1<q^{1/d}$, that the mean surface energy 
$\frac{\partial}{\partial\beta K}g_{\text{f}}$ 
is discontinuous near 
$\beta K=\ln \left(1+q^{1/(d-1)}\right)$.

Namely, let $\langle\cdot\rangle^{p}$ denote the infinite volume expectation
corresponding to the boundary condition $p$:%
\[
\langle f\rangle^{p}(\beta J,\beta K)=\lim_{L\rightarrow\infty,M\rightarrow
\infty}\frac{1}{Z^{p}(\Omega)}\sum_{\sigma_{\Omega}\in\mathcal{Q}^{\Omega}%
}fe^{-\beta H}\chi_{\Omega}^{p}%
\]
defined for local observable $f$ and let $e^{-\tau}$ be defined by (\ref{eq:3.23}%
) below. As a consequence of our main result (Theorem \ref{T:unicity} in
Section 3), we have the following

\begin{corollary}
Assume that $e^{\beta J}-1<q^{1/d}$ and $q$ is large enough, then there exists
a unique value $K_{t}$ such that

\begin{description}
\item[(i)] for any n.n. pair $ij$ of the surface%
\begin{alignat*}{3}
\langle\delta(\sigma_{i},\sigma_{j})\rangle^{\text{f}}(\beta J,\beta K)  &
\leq O(e^{-\tau})\quad &\text{for}\quad K\leq K_{t}\\
\langle\delta(\sigma_{i},\sigma_{j})\rangle^{\text{fo}}(\beta J,\beta K)  &
\geq1-O(e^{-\tau})\quad &\text{for}\quad K\geq K_{t}%
\end{alignat*}

\item[(ii)] for any n.n. pair $ij$ between the surface and the first layer
\begin{align*}
\langle\delta(\sigma_{i},\sigma_{j})\rangle^{\text{f}}(\beta J,\beta K)  &
\leq O(e^{-\tau})\;\\
\langle\delta(\sigma_{i},\sigma_{j})\rangle^{\text{fo}}(\beta J,\beta K)  &
\leq O(e^{-\tau})
\end{align*}

\end{description}
\end{corollary}

In that theorem the ratios of the partition functions entering in the
definition of the surface free energy $g_{\text{f}}$ (with both $Z^{\text{f}%
}(\Omega)$ and $Z^{\text{fo}}(\Omega)$) are expressed in terms of partition
functions of gas of polymers interacting through a two-body hard-core
exclusion potential. For $e^{\beta J}-1<q^{1/d}$ and $q$ large, the associated
activities   are small according the values of $K$ namely for $K\leq K_{t}$
with the free boundary condition and for $K\geq K_{t}$ with the free-ordered
boundary condition. The system is then controlled by convergent cluster expansion.

In the next subsection, we give the expression of the
Fortuin-Kasteleyn representation of the partition function
$Z^{p}(\Omega)$. We also present, in Theorems 2.1 and  2.2, the
high and low temperature expansions of the Fortuin-Kasteleyn
representation of the bulk partition function $Q(\Lambda)$ that
converge respectively for $e^{\beta J}-1<q^{1/d}$ and $e^{\beta
J}-1>q^{1/d}$ whenever $q$ is large enough.

\section{Random cluster models}

\setcounter{equation}{0}

\subsection{The Fortuin--Kasteleyn (FK) representation}

By using the expansion $e^{\beta K_{ij}\delta(\sigma_{i},\sigma_{j}%
)}=1+(e^{\beta K_{ij}}-1)\delta(\sigma_{i},\sigma_{j})$, we obtain the
Fortuin--Kasteleyn representation \cite{FK} of the partition function:
\begin{equation}
Z^{p}(\Omega)=\sum_{X\subset B(\Omega)}\prod_{\langle i,j\rangle\in
X}(e^{\beta K_{ij}}-1)q^{N_{\Omega}^{p}(X)} \label{eq:2.1}%
\end{equation}
where $B(\Omega)=\{\langle i,j\rangle:i\in\Omega,j\in\Omega\}$ is the set of
bonds with both endpoints belonging to $\Omega$, and $N_{\Omega}^{p}(X)$ is
the number of connected components (regarding an isolated site $i\in\Omega$ as
a component) of a given $X\subset B(\Omega)$. These numbers depend on the
considered boundary condition; introducing $S(X)$ as the set of sites that
belong to some bond of $X$ and $C(X|V)$ as the number of connected components
(single sites are not included) of $X$ that do not intersect the set of sites
$V$, they are given by:
\begin{align*}
&  N_{\Omega}^{\text{f}}(X)=|\Omega|-|S(X)|+C(X|\emptyset)\\
&  N_{\Omega}^{\text{o}}(X)=|\Omega|-|S(X)\cup\partial\Omega|+C(X|\partial
\Omega)\\
&  N_{\Omega}^{\text{fo}}(X)=|\Omega|-|S(X)\cup\partial_{s}\Omega
|+C(X|\partial_{s}\Omega)\\
&  N_{\Omega}^{\text{of}}(X)=|\Omega|-|S(X)\cup\partial_{b}\Omega
|+C(X|\partial_{b}\Omega)
\end{align*}
Hereafter $|E|$ denotes the number of elements of the set $E$. The Boltzmann
weight in (\ref{eq:2.1}) can be written in terms of the following
Hamiltonian:
\begin{equation}
H_{\Omega}^{p}(X)=-\beta_{s}|X_{s}|-\beta_{b}|X_{b}|-N_{\Omega}^{p}%
(X)+|\Omega| \label{eq:2.2}%
\end{equation}
where $X_{s}=X\cap B(\mathbb{L}_{0})$, $X_{b}=X\setminus X_{s}$, and%

\begin{equation}
\left\{
\begin{array}
[c]{rl}
& \beta_{s}\equiv\displaystyle\frac{\ln(e^{\beta K}-1)}{\ln q}\\
& \beta_{b}\equiv\displaystyle\frac{\ln(e^{\beta J}-1)}{\ln q}%
\end{array}
\right.  \label{eq:2.3}%
\end{equation}
Actually, one has:
\begin{equation}
Z^{p}(\Omega)=q^{|\Omega|}\sum_{X\subset B(\Omega)}q^{-H_{\Omega}^{p}(X)}
\label{eq:2.4}%
\end{equation}
Hence from this representation where the configurations are given by subsets
of bonds of the lattice, we are left to analyze a model with an inverse
temperature given by $\ln q$ and Hamiltonian given by (\ref{eq:2.2}) where
$\beta_{s}$ and $\beta_{b}$ are independent (real--valued) coupling constants.
The next subsection is devoted to the study of the ground states of this model.

\subsection{Ground states}

To study the diagram of ground states of the Fortuin--Kasteleyn
representation, it is convenient to introduce the formal Hamiltonian
\begin{equation}
H(X)\equiv-\beta_{s}|X_{s}|-\beta_{b}|X_{b}|+|S(X)|-C(X) \label{eq:2.5}%
\end{equation}
where $C(X)=C(X|\emptyset)$ is the number of connected components of $X$, and
the relative Hamiltonian
\begin{equation}
H(Y|X)\equiv H(Y)-H(X) \label{eq:2.6}%
\end{equation}
Note, that this last expression makes sense for any configurations $X$ and $Y
$ that coincide almost everywhere (a.e.), i.e. that differ only on a finite
set of bonds. A configuration $X$ is called ground state, if for any
configuration $Y=X$ $a.e.$, one has
\[
H(Y|X)\geq0
\]
For a fixed bond $b$ of the lattice let $n_{b}$ be the occupation number:
$n_{b}=1$ if $b$ belongs to $X$ and $0$ otherwise. The structure of ground
states invariant under horizontal translations is indicated in Fig. 1 (for
positive couplings $\beta_{s}$ and $\beta_{b}$).

\begin{center}
\setlength{\unitlength}{6.5mm} \begin{picture}(14,7) \put(1,-3.5){
\begin{picture}(0,0)
\drawline(0,0)(0,7)
\drawline(0,0)(8,0)       %
\put(7.8,-0.15){$\blacktriangleright$} \put(8,-0.7){$\beta_{b}$}
\put(-0.175,6.9){$\blacktriangle$} \put(-0.7,7){$\beta_{s}$}
\put(1.9,-0.7){$\frac{1}{d}$} \put(5.9,-0.7){$1$}
\put(-1.1,2.85){$\frac{1}{d-1}$}
\drawline(0,3)(2,3) \drawline(2,2)(6,0) \drawline(2,0)(2,6.5)
\put(.5,1.5){I} \put(.5,4.5){II} \put(4.2,3.7){IV}
\put(2.5,0.5){III} \put(2.3,2.1){$S_2$} \put(2.3,3.1){$S_1$}
\end{picture}
}
\end{picture}

\vspace{4cm}

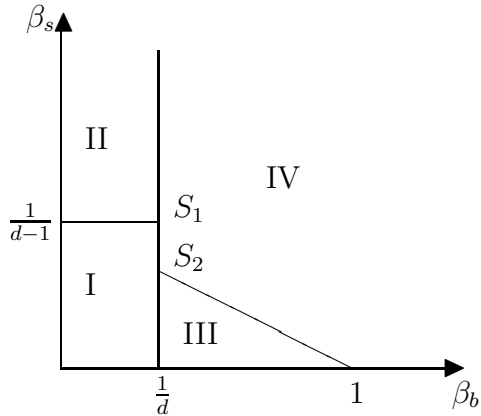
{\footnotesize {FIGURE 1 : The diagram of ground states.} }
\end{center}

There are four regions with a unique ground state:

\begin{enumerate}
\item In region I ( $\beta_{b}<\frac{1}{d}$ , $\beta_{s}<\frac{1}{d-1}$, the
ground state is the configuration $X^{\text{f}}=\emptyset$ .

\item In region II ($\beta_{b}<\frac{1}{d}$ , $\beta_{s}>\frac{1}{d-1}$, it is
the configuration $X^{\text{fo}}=B(\mathbb{L}_{o})$.

\item In region III ( $\beta_{b}>\frac{1}{d}$ , $\beta_{s}<\frac{1}%
{d-1}(1-\beta_{b})$, it is the configuration $X^{\text{of}}=B(\mathbb{L}%
\setminus\mathbb{L}_{o})$.

\item In region IV ( $\beta_{b}>\frac{1}{d}$ , $\beta_{s}>\frac{1}%
{d-1}(1-\beta_{b})$, it is the configuration $X^{\text{o}}=B(\mathbb{L})$.
\end{enumerate}

On the separating lines between any two regions the corresponding ground
states coexists. 
The g.s. 
$X^{\text{f}}$ 
coexists with 
$X^{\text{fo}}$
and
$X^{\text{o}}$ 
at the point $S_{1}$ 
and with 
$X^{\text{of}}$ 
and 
$X^{\text{o}}$ 
at the point $S_{2}$.

On the part of the line 
$\beta_{d}=1/d$ 
above and on $S_{1}$,
there are furthermore infinitely many g.s.\ that coexist with the
previous ones, namely all configurations say $X_{k}$, $k=2,3,...$,
with occupied bonds $n_{b}=1$ below and on the $k$ layer and empty
bonds $n_{b}=0$ otherwise. 
Analogously, on the part of the line
$\beta_{d}=1/d$ below and on $S_{2}$ there are also infinitely
many g.s.\ coexisting with the previous ones, namely all
configurations $X_{k}^{\prime}$, $k=2,3,...$, with occupied bonds
above and on the $k$ layer and empty bonds otherwise. 
Finally,
when $\beta_{s}=0$ and $\beta_{b}\geq1$, comes in addition to this
list the configuration 
$\tilde{X}^{\text{of}}=B(\mathbb{L})\setminus B(\mathbb{L}_{o})$.

This suggest that for large $q$ the extraordinary transitions will occur
through a sequence of layering transitions. Here we concentrate on the surface
transitions, but let us mention a particular study of layering transitions and
entropic repulsion for a simplified Solid-On-Solid type model in \cite{DM},
\cite{CF}.

\subsection{FK representation in the high temperature regime}

In terms of the Fortuin--Kasteleyn representation, the partition function
$Q^{\text{f}}(\Lambda)$ reads:
\begin{equation}
Q^{\text{f}}(\Lambda)=q^{|\Lambda|}\sum_{Y\subset B(\Lambda)}q^{\beta
_{b}|Y|-|S(Y)|+C(Y)} \label{eq:2.7}%
\end{equation}
We will first recall that this system is described by a gas of polymers
interacting through two-body hard-core potential and with activity given by
(\ref{eq:2.8}) below.

Indeed, first let us call polymer, a connected subset of
$B(\mathbb{L})$ (in the $\mathbb{R}^{d}$ sense). We will use
$\mathcal{P}(\Lambda)$ to denote the set of polymers whose bonds
belong to $B(\Lambda)$. Two polymers $\gamma_{1}$ and $\gamma_{2}$
are said compatible (we will write $\gamma_{1}\thicksim
\gamma_{2}$) if they do not intersect, i.e.\ if the set
$S(\gamma_{1})\cap S(\gamma_{2})$ is empty. They are said
incompatible otherwise and we will use the notation
$\gamma_{1}\nsim\gamma_{2}$. A family of polymers is said
compatible if any two polymers of the family are compatible and we
will use $\mathbf{P}(\Lambda)$ to denote the set of compatible
families of polymers $\gamma\in\mathcal{P}(\Lambda)$.

Clearly each subset $Y$ of $B(\Lambda)$ may be decomposed (in a unique way) in
a compatible family of polymers $Y=\{\gamma_{1},\ldots,\gamma_{n}\}$.
Introducing the activity of polymers by
\begin{equation}
\varphi_{\text{f}}(\gamma)=q^{\beta_{b}|\gamma|-|S(\gamma)|+1} \label{eq:2.8}%
\end{equation}
one has:
\begin{equation}
Q^{\text{f}}(\Lambda)=q^{|\Lambda|}\sum_{Y\in\mathbf{P}(\Lambda)}\prod
_{\gamma\in Y}\varphi_{f}(\gamma) \label{eq:2.9}%
\end{equation}
(with the sum running over compatible families of polymers including the
empty-set with weight equal to $1$).

We will now introduce multi-indexes in order to write the logarithm of this
partition function as a sum over these multi-indexes (see \cite{M}). A
multi-index $C$ is a function from the set $\mathcal{P}(\Lambda)$ into the set
of non negative integers, and we let $\text{supp}\,C=\left\{  \gamma
\in\mathcal{P}(\Lambda):C(\gamma)\geq1\right\}  $. We define the truncated
functional
\begin{equation}
\Phi_{0}(C)=\frac{a(C)}{\prod_{\gamma}C(\gamma)!}\prod_{\gamma}\varphi
_{f}(\gamma)^{C(\gamma)} \label{eq:2.10}%
\end{equation}
where the factor $a(C)$ is a combinatoric factor defined in terms of the
connectivity properties of the graph $G(C)$ with vertices corresponding to
$\gamma\in$ supp$\,C$ (there are $C(\gamma)$ vertices for each $\gamma\in$
supp$\,C$) that are connected by an edge whenever the corresponding polymers
are incompatible). Namely, $a(C)=0$ and hence $\Phi_{0}(C)=0$ unless $G(C)$ is
a connected graph in which case $C$ is called a \emph{cluster} and
\begin{equation}
a(C)=\sum_{G\subset G(C)}(-1)^{\left\vert e(G)\right\vert } \label{eq:2.11}%
\end{equation}
Here the sum goes over connected subgraphs $G$ whose vertices coincide with
the vertices of $G(C)$ and $\left\vert e(G)\right\vert $ is the number of
edges of the graph $G$. If the cluster $C$ contains only one polymer, then
$a(\gamma)=1$. In other words, the set of all bonds of polymers belonging to a
cluster $C$ is connected. The support of a cluster is thus a polymer and it is
then convenient to define the following new truncated functional
\begin{equation}
\Phi(\gamma)=\sum_{C:\text{supp}\,C=\gamma}\Phi_{0}(C) \label{eq:2.12}%
\end{equation}
As a result we have

\begin{theorem}
\label{CE1} Assume that $\beta_{b}<1/d$ and $c_{0}\nu_{d}q^{-\frac{1}{d}%
+\beta_{b}}\leq1$, where $\nu_{d}=(2d)^{2}$ and $c_{0}=(\sqrt{3}+2)e^{\frac
{2}{\sqrt{3}+1}}$, then
\begin{equation}
Q^{\text{f}}(\Lambda)=q^{|\Lambda|}\exp\left\{  \sum_{\gamma\in\mathcal{P}%
(\Lambda)}\Phi(\gamma)\right\}  \label{eq:2.13}%
\end{equation}
with a sum running over (non-empty) polymers of $\Lambda$, and the truncated
functional $\Phi$ satisfies the estimates%
\begin{equation}
\left\vert \Phi(\gamma)\right\vert \leq\left\vert \gamma\right\vert \left(
c_{0}\nu_{d}q^{-\frac{1}{d}+\beta_{b}}\right)  ^{\left\vert \gamma\right\vert
} \label{eq:2.14}%
\end{equation}

\end{theorem}

\begin{proof}
We first notice that easy geometrical observations \cite{LMeR,LMMRS} gives the
inequality $\frac{|\gamma|}{d}-|S(\gamma)|+1\leq0$ which in turn implies:
\begin{equation}
\varphi_{\text{f}}(\gamma)\leq q^{-(\frac{1}{d}-\beta_{b})|\gamma|}
\label{eq:2.15}%
\end{equation}
On the other hand the number of polymers $\gamma$ of length $|\gamma|=n$
containing a given site is less than $(\nu_{d})^{n}$. This follows from the
fact that for any polymer there exists a path going through the center of
every bond at most twice (see \cite{O}). Then we have $\sum_{\gamma\ni
x}\varphi_{\text{f}}(\gamma)<\infty$ as soon as $\nu_{d}q^{-\frac{1}{d}%
+\beta_{b}}<1$.

According ref. \cite{M}, under the condition
\begin{equation}
\varphi_{\text{f}}(\gamma_{0})\leq\left(  e^{\mu(\gamma_{0})}-1\right)
\exp\left[  -\sum_{\gamma\nsim\gamma_{0}}\mu(\gamma)\right]  \label{eq:2.16}%
\end{equation}
where $\mu$ is a positive function, one has%
\begin{equation}
\ln\sum_{Y\in\mathbf{P}(\Lambda)}\prod_{\gamma\in Y}\varphi_{\text{f}}%
(\gamma)=\sum_{C}\Phi_{0}(C) \label{eq:2.17}%
\end{equation}
where the last sum runs over (non empty) clusters of polymers in
$\mathcal{P}(\Lambda)$, and the truncated functional $\Phi_{0}$ satisfies the
estimate%
\begin{equation}
\sum_{C:C(\gamma_{1})\geq1}|\Phi_{0}(C)|\leq\mu(\gamma_{1}) \label{eq:2.18}%
\end{equation}
We choose $\mu(\gamma)=\left(  a\nu_{d}\right)  ^{-\left\vert \gamma
\right\vert }$ to get by taking into account the above remark on the entropy
of polymers
\begin{equation}
\sum_{\gamma\nsim\gamma_{0}}\mu(\gamma)\leq|S(\gamma_{0})|\sum_{n=1}^{\infty
}a^{-n}\leq(|\gamma_{0}|+1)\frac{a^{-1}}{1-a^{-1}}\leq\frac{2}{a-1}|\gamma
_{0}| \label{eq:2.19}%
\end{equation}
for $a>1$. Since $\mu(\gamma_{0})\leq e^{\mu(\gamma_{0})}-1$, the bound
(\ref{eq:2.15}) gives that the condition (\ref{eq:2.16}) will be satisfied
whenever
\begin{equation}
\nu_{d}q^{-\frac{1}{d}+\beta_{b}}ae^{\frac{2}{a-1}}\leq1 \label{eq:2.20}%
\end{equation}
The value $a_{0}=\sqrt{3}+2$ that minimizes the function $ae^{\frac{2}{a-1}}$
provides the conditions given in the theorem. By definition (\ref{eq:2.12}),
the relation (\ref{eq:2.17}) yields (\ref{eq:2.13}) while the relations
(\ref{eq:2.18}) and (\ref{eq:2.19}) yield:
\begin{align}
\left\vert \Phi(\gamma_{0})\right\vert  &  \leq e^{-\lambda\left\vert
\gamma_{0}\right\vert }\sum_{C:\text{supp}\,C=\gamma_{0}}|\Phi_{0}%
(C)|\prod_{\gamma}e^{\lambda C(\gamma)}\label{eq:2.21}\\
&  \leq e^{-\lambda\left\vert \gamma_{0}\right\vert }\sum_{\gamma\nsim
\gamma_{0}}\sum_{C:C(\gamma)\geq1}|\Phi_{0}(C)|\prod_{\gamma}e^{\lambda
C(\gamma)}\leq\frac{2}{a_{0}-1}\left\vert \gamma_{0}\right\vert e^{-\lambda
\left\vert \gamma_{0}\right\vert }\nonumber
\end{align}
provided $\nu_{d}e^{\lambda}q^{-\frac{1}{d}+\beta_{b}}a_{0}e^{\frac{2}%
{a_{0}-1}}\leq1$; we choose for $\lambda$ the value that realizes the equality
to get (by taking into account that $\frac{2}{a_{0}-1}\leq1$) the estimate
(\ref{eq:2.14}).
\end{proof}

Let us mention that a recent study of these expansions for the Potts model on
a general graph, i.e.\ not restricted to nearest neighbor interactions, is
given in ref.\ \cite{So}.

\subsection{FK representation in the low temperature regime}

We now derive an expansion of the bulk partition function $Q^{\text{o}%
}(\Lambda)$ at \textquotedblleft temperature\textquotedblright\ $\beta
_{b}>\frac{1}{d}$. The FK representation of this partition function reads%
\begin{equation}
Q^{\text{o}}(\Lambda)=\sum_{Y\subset B(\Lambda)}q^{\beta_{b}|Y|
+N_{\Lambda}^{\text{o}}(Y)
}
=q^{\beta_{b}|B(\Lambda)|}\sum_{Y\subset B(\Lambda)}q^{-\beta
_{b}|B(\Lambda)\backslash Y|+N_{\Lambda}^{\text{o}}(Y)}%
\end{equation}
where $N_{\Lambda}^{\text{o}}(Y)=\left\vert \Lambda\right\vert -\left\vert
S(X)\cup\partial\Lambda\right\vert +C(X|\partial\Lambda)$. The expansion is
mainly based on a duality property and we first recall geometrical results on
Poincar\'{e} and Alexander duality (see e.g. \cite{L},\cite{Al},\cite{DW}%
,\cite{KLMR}).

We first consider the lattice $\mathbb{Z}^{d}$ and the associated cell-complex
$\mathbf{L}$ whose objects $s_{p}$ are called $p$-cells ($0\leq p\leq d$):
$0$-cells are vertices, $1$-cells are bonds, $2$-cells are plaquettes etc...:
a $p$-cell may be represented as $(x;\sigma_{1}e_{1},...,\sigma_{p}e_{p})$
where $x\in\mathbb{Z}^{d},(e_{1},...,e_{d})$ is an orthonormal base of
$\mathbb{R}^{d}$ and $\sigma_{\alpha}=\pm1,\alpha=1,...,d $. Consider also the
dual lattice
\[
(\mathbb{Z}^{d})^{\ast}=\left\{  x=(x_{1}+\frac{1}{2},...,x_{d}+\frac{1}%
{2}):x_{\alpha}\in\mathbb{Z},\alpha=1,...,d\right\}
\]
and the associated cell complex $\mathbf{L}^{\ast}$. There is a one to-one
correspondence
\begin{equation}
s_{p}\leftrightarrow s_{d-p}^{\ast} \label{eq:2.23}%
\end{equation}
between $p$-cells of the complex $\mathbf{L}$ and the $d-p$-cells of
$\mathbf{L}^{\ast}$. In particular to each bond $s_{1}$ corresponds the
hypercube $s_{d-1}^{\ast}$ that crosses $s_{1}$ in its middle. The dual
$E^{\ast}$ of a subset $E\subset\mathbf{L}$ is the subset of elements of
$L^{\ast}$ that are in the one-to-one correspondence (\ref{eq:2.23}) with the
elements of $E$.

We now turn to the Alexander duality in the particular case under
consideration in this paper. Let $Y\subset B(\Lambda)$ be a set of bonds. We
define the A-dual of $Y$ as
\begin{equation}
\widehat{Y}=\left(  B(\Lambda)\setminus Y\right)  ^{\ast}%
\end{equation}
As a property of Alexander duality one has%
\begin{align}
\left\vert \widehat{Y}\right\vert  &  =\left\vert B(\Lambda)
\setminus Y\right\vert \\
N_{\Lambda}^{\text{o}}(Y)  &  =N_{\text{cl}}(\widehat{Y})
\end{align}
where $N_{\text{cl}}(\widehat{Y})$ denotes the number of independent closed
($d-1$)-surfaces of $\widehat{Y}$. We thus get%
\begin{equation}
Q^{\text{o}}(\Lambda)=q^{\beta_{b}|B(\Lambda)|}\sum_{\widehat{Y}\subset\left[
B(\Lambda)\right]  ^{\ast}}q^{-\beta_{b}|\widehat{Y}|+N_{\text{cl}}%
(\widehat{Y})}%
\end{equation}

Now as in the previous subsection, the system can be described by a gas of
polymers interacting through hard core exclusion potential. Indeed, we
introduce (hat)-polymers as connected subsets (in the $\mathbb{R}^{d}$ sense)
of $(d-1)$-cells of $\mathbf{L}^{\ast}$ and let $\widehat{\mathcal{P}}%
(\Lambda)$ denote the set of polymers whose $(d-1)$-cells belong to $\left[
B(\Lambda)\right]  ^{\ast}$. Two polymers $\widehat{\gamma}_{1}$and
$\widehat{\gamma}_{2}$ are compatible (we will write $\widehat{\gamma}%
_{1}\thicksim\widehat{\gamma}_{2}$) if they do not intersect and incompatible
otherwise (we will write $\widehat{\gamma}_{1}\nsim\widehat{\gamma}_{2}$). A
family of polymers is said compatible if any two polymers of the family are
compatible and we will use $\widehat{\mathbf{P}}(\Lambda)$ to denote the set
of compatible families of polymers $\widehat{\gamma}\in\mathcal{P}(\Lambda)$.
Introducing the activity of polymers by
\begin{equation}
\varphi_{\text{o}}(\widehat{\gamma})=q^{-\beta_{b}|\widehat{\gamma
}|+N_{\text{cl}}(\widehat{\gamma})}%
\end{equation}
one has:
\begin{equation}
Q^{\text{o}}(\Lambda)=q^{\beta_{b}|B(\Lambda)|}\sum_{\widehat{Y}\in
\widehat{\mathbf{P}}(\Lambda)}\prod_{\widehat{\gamma}\in\widehat{Y}}%
\varphi_{\text{o}}(\widehat{\gamma})
\end{equation}
(with the sum running over compatible families of polymers including the
empty-set with weight equal to $1$).\ As in the previous subsection, we
introduce multi-indexes $\widehat{C}$ as functions from the set $\widehat
{\mathcal{P}}(\Lambda)$ into the set of non negative integers, and the
truncated functional
\begin{align}
\widehat{\Phi}_{0}(\widehat{C})  &  
=\frac{a(\widehat{C})}{\prod_{\widehat{\gamma}}\widehat{C}(\widehat{\gamma})!}
\prod_{\widehat{\gamma}}\varphi_{\text{o}%
}(\widehat{\gamma})^{C(\widehat{\gamma})}\\
\widehat{\Phi}(\widehat{\gamma})  &  =\sum_{\widehat{C}:\text{supp}%
\,\widehat{C}=\widehat{\gamma}}\widehat{\Phi}_{0}(\widehat{C})
\end{align}
Here $\text{supp}\,\widehat{C}=\left\{  \widehat{\gamma}\in\widehat
{\mathcal{P}}(\Lambda):C(\widehat{\gamma})\geq1\right\}  $ and $a(\widehat
{C})$ is defined by (\ref{eq:2.11}) with a graph $G(\widehat{C})$ with
vertices corresponding to $\widehat{\gamma}\in\text{supp}\,\widehat{C}$ that
are connected by an edge when the corresponding polymers are incompatible.
Again, $\widehat{\Phi}_{0}(\widehat{C})=0$ if $\,\widehat{C}$ is not a cluster
i.e.\ if the set of all $d-1$-cells of polymers belonging to $\text{supp}%
\,\widehat{C}$ is not connected.

\begin{theorem}
Assume that $\beta_{b}>1/d$ and $\widehat{c}_{0}\widehat{\nu}_{d}q^{-\beta
_{b}+\frac{1}{d}}\leq1$, where $\widehat{\nu}_{d}=d^{2}2^{4(d-1)}$, and
$\widehat{c}_{0}=\left[  1+2^{d-2}(1+\sqrt{1+2^{3-d}})\right]  \exp\left[
\frac{2}{1+\sqrt{1+2^{3-d}}}\right]  $, then
\begin{equation}
Q^{\text{o}}(\Lambda)=e^{\beta_{b}|B(\Lambda)|}\exp\left\{  \sum
_{\widehat{\gamma}\in\widehat{\mathcal{P}}(\Lambda)}\widehat{\Phi}%
(\widehat{\gamma})\right\}
\end{equation}
with a sum running over (non-empty) polymers, and the truncated functional
$\widehat{\Phi}$ satisfies the estimates%
\begin{equation}
\left\vert \widehat{\Phi}(\widehat{\gamma})\right\vert \leq\left\vert
\widehat{\gamma}\right\vert \left(  \widehat{c}_{0}\widehat{\nu}_{d}%
q^{-\beta_{b}+\frac{1}{d}}\right)  ^{\left\vert \widehat{\gamma}\right\vert }%
\end{equation}

\end{theorem}

\begin{proof}
We first notice that the obvious geometrical inequality $N_{\text{cl}%
}(\widehat{\gamma})\leq|\widehat{\gamma}|/d$ implies:
\begin{equation}
\varphi_{\text{o}}(\widehat{\gamma})\leq q^{-(\beta_{b}-\frac{1}{d}%
)|\widehat{\gamma}|}%
\end{equation}
On the other hand the number of polymers $\widehat{\gamma}$ of length
$|\widehat{\gamma}|=n$ containing a given vertex is less than $(\widehat{\nu
}_{d})^{n}$. To see it, we first observe that the number of $p$-cells that
share a same vertex equals $2^{p}\binom{d}{p}$: $2^{p}$ is the choice for the
signs of the $\sigma$'s and the binomial coefficient $\binom{d}{p}$ is the
choice for $p$ vectors $e_{\alpha}$ among $(e_{1},...,e_{d})$. This implies
(by duality) that a $(d-1)$-cell contains $2^{d-1}$ vertices and hence that
the number of $(d-1)$-cells connected (i.e. sharing at least a vertex) with a
given $(d-1)$-cell is less than $2^{(d-1)}(d2^{(d-1)}-1)$. Finally, one uses
the fact that for any polymer of length $n$ there exists a path of length less
than $2n$ going through the center of every $(d-1)$-cell.

For a function $\mu(\widehat{\gamma})=\left(  a\widehat{\nu}_{d}\right)
^{-\left|  \widehat{\gamma}\right|  }$ we get
\begin{equation}
\sum_{\widehat{\gamma}\nsim\widehat{\gamma}_{0}}\mu(\widehat{\gamma})\leq
\sum_{n=1}^{\infty}|S(\widehat{\gamma}_{0})|a^{-n}\leq\frac{2^{d-1}}%
{a-1}|\widehat{\gamma}_{0}|
\end{equation}
if $a>1$ and the convergence condition (\ref{eq:2.15}) for the case under
consideration will be satisfied whenever
\begin{equation}
\widehat{\nu}_{d}q^{-\beta_{b}+\frac{1}{d}}ae^{\frac{2^{d-1}}{a-1}}\leq1
\end{equation}
Here we choose $a_{0}=1+2^{d-2}(1+\sqrt{1+2^{3-d}})$ that minimizes the
function $ae^{\frac{2^{d-1}}{a-1}}$.

The remainder of the proof is analog to that of Theorem \ref{CE1}
\end{proof}

\section{Surface transition in the high temperature bulk regime}

\setcounter{equation}{0}

\subsection{Hydra model}

To study the surface free energy $g_{\text{f}}$ we shall express the ratio of
partition functions entering in its definition (\ref{eq:1.4}) in terms of a
partition function of geometrical objects to be called \emph{hydras}.

\begin{definition}
A connected set of bonds $\delta\subset B(\Omega)$ is called \emph{hydra} if
there exists a bond of $\delta$ which has one of its endpoint on the boundary
surface $\Sigma$ (i.e. if $S(\delta)\cap\Sigma\neq\emptyset$).

Any bond of the hydra with one endpoint in the boundary surface $\Sigma$ and
one endpoint in the bulk $\Lambda$ is called \emph{neck} of the hydra.

A connected component of bonds of the hydra with two endpoints in the boundary
surface is called \emph{foot} of the hydra.

A connected component of bonds of the hydra with two endpoints in the bulk is
called \emph{head} of the hydra (heads of hydras are polymers introduced in
Subsection 2.3).
\end{definition}

\begin{center}
\setlength{\unitlength}{8 mm} \begin{picture}(17,6)(0,0)
\put(3,0){
\begin{picture}(0,0)
\drawline(0,0)(3,0) \drawline(5,0)(8,0)
\dottedline{.1}(1,0)(1,1) \dottedline{.1}(2,0)(2,1)
\dottedline{.1}(6,0)(6,1) \dottedline{.1}(8,0)(8,1)
\dottedline{.1}(10,0)(10,1)
\dashline{.1}(0,1)(1,1) \dashline{.1}(0,2)(1,2)
\dashline{.1}(1,1)(1,2) \dashline{.1}(0,1)(0,2)
\dashline{.1}(1,2)(1,3)
\dashline{.1}(2,1)(2,2) \dashline{.1}(2,1)(2,2)
\dashline{.1}(2,2)(2,3) \dashline{.1}(2,3)(4,3)
\dashline{.1}(4,2)(4,3)
\dashline{.1}(2,2)(6,2) \dashline{.1}(6,1)(6,2)
\dashline{.1}(5,2)(5,3) \dashline{.1}(5,3)(6,3)
\dashline{.1}(6,3)(6,4)
\dashline{.1}(8,1)(10,1)\dashline{.1}(9,1)(9,3)
\dashline{.1}(9,3)(11,3)\dashline{.1}(9,2)(10,2)
\dashline{.1}(10,2)(10,3)
\end{picture}
}
\end{picture}

\vspace{1cm} {\footnotesize {FIGURE 2 : A hydra, in two dimensions (a
dimension not considered in this paper), with $2$ feet (components of full
lines), $5$ necks (dotted lines), and $3$ heads (components of dashed lines).}
}
\end{center}

We let $\mathcal{H}(\Omega)$ denote the set of hydras of $\Omega$. Two hydras
$\delta_{1}$ and $\delta_{2}$ are said compatible (we will write $\delta
_{1}\thicksim\delta_{2}$) if $S(\delta_{1})\cap S(\delta_{2})\neq\emptyset$. A
family of hydras is said compatible if any two hydras of the family are
compatible and we let $\mathbf{H}(\Omega)$ denote the set of compatible
families of hydras $\delta\in\mathcal{H}(\Omega) $.

Clearly, a connected subset of bonds included in $B(\Omega)$ is either a hydra
$\delta\in\mathcal{H}(\Omega)$ or a polymer $\gamma\in\mathcal{P}(\Lambda)$
(defined in subsection 2.3) and thus any subset of $B\in B(\Omega)$ is a
disjoint union of a compatible family of hydras $X\in\mathbf{H}(\Omega)$ with
a compatible family of polymers $Y\in\mathbf{P}(\Lambda)$.

Then, the partition function $Z^{p}(\Omega)$ defined by (\ref{eq:2.4}) reads
\begin{equation}
Z^{p}(\Omega)=q^{\left\vert \Omega\right\vert }\sum_{X\in\mathbf{H}(\Omega
)}q^{-H_{\Omega}^{p}(X)}\sum_{Y\in\mathbf{P}(\Lambda):Y\thicksim X}%
\prod_{\gamma\in Y}\varphi_{\text{f}}(\gamma) \label{eq:3.1}%
\end{equation}
where we take into account that the Boltzmann weight $q^{-H_{\Omega}^{p}(Y)}$
of the subsets $Y\subset B(\Lambda)$ coincides with the bulk one given in
(\ref{eq:2.7}) and the compatibility $Y\thicksim X$ means $S(Y)\cap
S(X)=\emptyset$.

According to Subsection 2.3, the last sum in the RHS of the above formula can
be exponentiated as: $\exp\left\{  \sum_{\gamma\in\mathcal{P}(\Lambda
);\gamma\thicksim X}\Phi(\gamma)\right\}  $. Hence dividing the above
partition function by the partition function $Q^{\text{f}}(\Lambda)$ we get by
taking into account Theorem~\ref{CE1}:%
\begin{align}
\Xi^{p}(\Omega)  &  \equiv\frac{Z^{p}(\Omega)}{Q^{\text{f}}(\Lambda
)}\label{eq:3.2}\\
&  =q^{\left\vert \Sigma\right\vert }\sum_{X\in\mathbf{H}(\Omega
)}q^{-H_{\Omega}^{p}(X)}\exp\left\{  -\sum_{\gamma\in\mathcal{P}%
(\Lambda);\gamma\nsim X}\Phi(\gamma)\right\} \nonumber
\end{align}
Hereafter the incompatibility $\gamma\nsim X$ means $S(\gamma)\cap
S(X)\neq\emptyset.$

$\Xi^{p}(\Omega)$ is thus the partition function of a gas of
hydras $X=\{\delta_{1},\ldots,\delta_{n}\}$ whose activities are
defined from the Hamiltonian $H_{\Omega}^{p}(X)$; they interact
through hard-core exclusion potential and through a long range
interaction potential (decaying exponentially in the distance
under the hypothesis of  Theorem~\ref{CE1}) defined on the
polymers of the bulk.

Notice that if we neglect this long range potential, the system of
hydras will reduce itself to a $(d-1)$ Potts model (in the FK
representation), when we moreover restrict to consider only hydras
without neck and without head. As well known, the $(d-1)$ Potts
model undergoes a temperature driven first order phase transition
(at some inverse temperature $\approx(d-1)^{-1}\ln q$), whenever q
is large enough and $d\geq3$. We will show that it is also the
case for the hydra model (\ref{eq:3.2}) implementing the fact that
the necks and the heads of hydras modify only weakly their
activities and that the long range interaction potential decays
exponentially (the needed assumptions are close to those of
Theorem~\ref{CE1}).

To this end it is convenient to first rewrite this potential in terms of a
model of \emph{aggregates}, a technique already developed in many analog
situations (see e.g. \cite{HKZ},\cite{DKS}). Let us introduce the
(real-valued) functional
\begin{equation}
\Psi(\gamma)=e^{-\Phi(\gamma)}-1 \label{eq:3.3}%
\end{equation}
defined on polymers $\gamma\in \mathcal{P}(\Lambda)$. An aggregate
$A$ is a family of polymers whose support,
$\text{supp}\,A=\cup_{\gamma\in A}\gamma$, is connected. Two
aggregates $A_{1}$ and $A_{2}$ are said compatible if
$\text{supp}\,A_{1}\cap\text{supp}\,A_{2}=\emptyset$. A family of
aggregates is said compatible if any two aggregates of the family
are compatible and we will use $\mathbf{A}(\Lambda)$ to denote the
set of compatible families of
aggregates. Introducing the statistical weight of aggregates by%
\begin{equation}
\omega(A)=\prod_{\gamma\in A}\Psi(\gamma) \label{eq:3.4}%
\end{equation}
we then get:%
\begin{align}
\exp\left\{  -\sum_{\substack{\gamma\in\mathcal{P}(\Lambda); \\\gamma\nsim X
}}\Phi(\gamma)\right\}   &  =\prod_{\substack{\gamma\in\mathcal{P}(\Lambda);
\\\gamma\nsim X}}(1+\Psi(\gamma))\nonumber\\
&  =\sum_{\mathcal{A}\in\mathbf{A}(\Lambda)}\prod_{\substack{A\in\mathcal{A};
\\A\nsim X}}\omega(A) \label{eq:3.5}%
\end{align}
where\ $A\nsim X$ means that every polymer of the aggregate $A$ is
incompatible with $X$. Since the support of aggregates is a
connected set of bonds i.e.\ a polymer, it is convenient (as it
was done for clusters in Subsection 2.3) to sum the statistical
weights (\ref{eq:3.4}) over aggregates
with same support. We thus define the weight%
\begin{equation}
\psi(\gamma)\equiv\sum_{A:\text{supp}\,A=\gamma}\omega(A) \label{eq:3.6}%
\end{equation}
with $A\nsim X$, to get%
\begin{equation}
\Xi^{p}\left(  \Omega\right)  =q^{\left\vert \Sigma\right\vert }\sum
_{X\in\mathbf{H}(\Omega)}q^{-H_{\Omega}^{p}(X)}\sum_{Y\mathbf{\in P}(\Lambda
)}\prod_{\substack{\gamma\in Y \\\gamma\nsim X}}\psi(\gamma) \label{eq:3.7}%
\end{equation}

The system is thus described by two families: a compatible family of hydras
and a compatible family of polymers each of these polymers being incompatible
with the family of hydras.

The next subsections are devoted to the large $q$ analysis of this model in
the frame of Pirogov-Sinai theory. As well known in this theory, it is useful
to introduce \emph{diluted} and \emph{crystal partition functions} for which
an important set of recurrence relation holds (see Lemma 3.2 below). Let us
introduce the diluted partition functions for our system .

We define the \emph{envelope} $E(M)$ of a set of sites of the lattice
$\mathbb{L}$ as the set of bonds of $B(\mathbb{L})$ that contain an endpoint
in $M$. Next it is convenient to define local observables with which the main
terms of the Hamiltonian (the number of bonds and the number of sites) can be written.

Let the energy per site of a configuration $X\subset B(\mathbb{L})$ be:%
\begin{equation}
e_{i}(X)=-\frac{\beta_{s}}{2}\left\vert E(i)\cap X_{s}\right\vert -\frac
{\beta_{b}}{2}\left\vert E(i)\cap X_{b}\right\vert +\left\vert i\cap
S(X)\right\vert \label{eq:3.8}%
\end{equation}
where $X_{s}=X\cap B(\mathbb{L}_{0}\mathbb{)}$ and $X_{b}=X\setminus X_{s}$.
We define, for (any) volume $\Omega\subset\mathbb{L}$, the diluted Hamiltonian
of a configuration $X=X^{p}$ a.e., as:%
\begin{equation}
H_{\Omega}^{\text{dil}}(X)=\sum_{i\in\Omega}e_{i}(X)-C(X)+C(X^{p})
\label{eq:3.9}%
\end{equation}
where the number of connected components $C(X^{\text{f}})$ of the ground state
$X^{\text{f}}$ equals $0$, and the number of connected component
$C(X^{\text{fo}})$ of the ground state $X^{\text{fo}}$ equals $1$.

Note that even though our model is defined on a $d$--dimensional box $\Omega$
it has a $(d-1)$-dimensional structure and the highest order\ of the logarithm
of partition functions behaves like $O(\left\vert \Sigma\right\vert )$. It
will be convenient to consider $\Omega$ as a set of lines. We let a line
$L(x)$ be a cylinder set of sites of $\mathbb{L}$ whose projection on the
boundary layer is the site $x$ and whose height is \ less than a given number
$M$: $L(x)=\{i\in\mathbb{L}\ |(i_{1},...,i_{d-1})=x\in\mathbb{L}_{0},i_{d}\leq
M\}$. We let $\mathbb{L}_{M}$ be the set of all such lines. For $\Omega
\subset\mathbb{L}_{M}$, we let $\Sigma=\Omega\cap\mathbb{L}_{0}$, be its
projection on the boundary layer, $\Lambda=\Omega\setminus\Sigma$ and
$\left\Vert \Omega\right\Vert =\left\vert \Sigma\right\vert $ be the number of
lines of $\Omega$. We will use $\mathbf{H}^{p}(\Omega)$ to denote the set of
compatible families of hydras defined on the envelope $E(\Omega)$ that
coincide with the ground state $X^{p}$ on the envelope $E(\partial\Omega)$ and
on the envelope $\ E(\mathbb{L}_{M}\setminus\Omega)$, and use $\mathbf{P}%
^{\text{dil}}\left(  \Lambda\right)  $ to denote the compatible families of
polymers defined on $E(\Omega)\setminus(E(\Sigma)\cup E(\partial\Omega))$.

The diluted partition function is defined by
\begin{equation}
\Xi_{p}^{\text{dil}}\left(  \Omega\right)  =\sum_{X\in\mathbf{H}^{p}(\Omega
)}q^{-H_{\Omega}^{\text{dil}}(X)}\sum_{Y\mathbf{\in P}^{\text{dil}}\left(
\Lambda\right)  }\prod_{\substack{\gamma\in Y \\\gamma\nsim X}}\psi(\gamma)
\end{equation}
Notice that the diluted Hamiltonian on ground states reads on set of lines
$\Omega\subset\mathbb{L}_{M}$:%
\begin{equation}
H_{\Omega}^{\text{dil}}(X^{p})=e_{p}\left\Vert \Omega\right\Vert
\end{equation}
where%
\begin{equation}%
\begin{array}
[c]{ll}%
e_{\text{f}} & =0\\
e_{\text{fo}} & =1-(d-1)\beta_{s}%
\end{array}
\label{eq:3.12}%
\end{equation}
Up to a boundary term $O(\partial\Sigma)$ one has $\ln\Xi^{p}\left(
\Omega\right)  =\left\Vert \Omega\right\Vert \ln q+\ln\Xi_{p}^{\text{dil}%
}\left(  \Omega\right)  $ so that
\begin{equation}
-\lim_{\Omega\uparrow\mathbb{L}}\frac{1}{\left\Vert \Omega\right\Vert }\ln
\Xi_{p}^{\text{dil}}\left(  \Omega\right)  =g_{\text{f}}+\ln q
\end{equation}
where $\Omega\uparrow\mathbb{L}$ means that we take first the limit
$M\rightarrow\mathbb{\infty}$ and then the limit $\Sigma\uparrow\mathbb{L}_{0}
$ in the van-Hove or Fisher sense.

Notice also the following bound%
\begin{equation}
0\leq\frac{\partial}{\partial\beta_{s}}\ln\Xi_{p}^{\text{dil}}\left(
\Omega\right)  \leq\left\Vert \Omega\right\Vert (d-1)\ln q \label{eq:3.14}%
\end{equation}

\subsection{contours}

Let $\Omega\subset\mathbb{L}_{M}$ and $(X,Y)$ be a configuration of our system
in $\Omega$: $X\in\mathbf{H}^{p}(\Omega),Y\in\mathbf{P}^{\text{dil}}%
(\Omega),Y\nsim X.$ A site $i\in\Omega$ is called \emph{p-correct}, if $X$
coincides with the ground state $X^{p}$ on the bonds of the envelope $E(i)$
and $Y$ does not contain any bond of $E(i)$. A line is called p-correct if all
the sites of the line are p-correct. In other words a line $L$ is f-correct if
$X\cap E(L)=\emptyset$ and $Y\cap E(L)=\emptyset$; a line $L$ is fo-correct if
$X\cap E(L)=B(\mathbb{L}_{0})\cap E(L)$ and $Y\cap E(L)=\emptyset$. Sites and
lines that are not p-correct are called \emph{incorrect}.

The set of incorrect lines of a configuration $(X,Y)$ is called
\emph{boundary} of the configuration $(X,Y)$.

A triplet $\Gamma=\{\text{supp}\, \Gamma,X(\Gamma),Y(\Gamma)\}$, where
$\text{supp}\, \Gamma$ is a maximal connected subset of the boundary of the
configuration $(X,Y)$ called support of $\Gamma$, $X(\Gamma)$ the restriction
of $X$ to the envelope $E(\text{supp}\, \Gamma)$ of the support of $\Gamma$,
and $Y(\Gamma)$ the restriction of $Y$ to $E(\text{supp}\, \Gamma)$, is called
\emph{contour} of the configuration $(X,Y)$. Hereafter a set of sites is
called connected if the graph that joins all the sites $i,j$ of this set with
$d(i,j)\leq1$ is connected.

A triplet $\Gamma=\{\text{supp}\,\Gamma,X(\Gamma),Y(\Gamma)\}$, where
$\text{supp}\,\Gamma$ is a connected set of lines is called \emph{contour} if
there exists a configuration $(X,Y)$\ such that $\Gamma$ is a contour of
$(X,Y)$. We will use $\left\vert \Gamma\right\vert $ to denote the number of
incorrect points of $\text{supp}\,\Gamma$ and $\left\Vert \Gamma\right\Vert $
to denote the number of lines of $\text{supp}\,\Gamma$ .

Consider the configuration having $\Gamma$ as unique contour; it
will be denoted $(X^{\Gamma},Y^{\Gamma})$. Let $L_{p}(\Gamma)$ be
the set of p-correct lines of
$\mathbb{L}_{M}\setminus\text{supp}\,\Gamma$. Obviously, either a
component of $L_{\text{f}}(\Gamma)$ is infinite or a component of
$L_{\text{fo}}(\Gamma)$ is infinite. In the first case $\Gamma$ is
called contour of the free class or f-contour and in the second
case it is called fo-contour. When $\Gamma$ is a p-contour (we
will let $\Gamma^{p}$ denote such contours) we use $\text{Ext}\
\Gamma$ to denote the unique infinite component of
$L_{p}(\Gamma)$; this component is called \emph{exterior} of the
contour.\ The set of remaining components of $L_{p}(\Gamma)$ is
denoted $\text{Int}_{p}\Gamma$ and the set $L_{m\neq p}(\Gamma)$
is denoted
$\text{Int}_{m}\Gamma$. The union $\text{Int}\Gamma=\text{Int}_{\text{f}%
}\Gamma\cup\text{Int}_{\text{fo}}\Gamma$ is called \emph{interior} of the
contour and $V(\Gamma)=\text{supp}\,\Gamma\cup\text{Int}\Gamma$.

Two contours $\Gamma_{1}$ and $\Gamma_{2}$ are said compatible if the union of
their supports is not connected. They are mutually compatible external
contours if $V(\Gamma_{1})\subset\text{Ext }\Gamma_{2}$ and $V(\Gamma
_{2})\subset\text{Ext }\Gamma_{1}$.

We will use $G(\Gamma^{p})$ to denote the set of configurations having
$\Gamma^{p}$ as unique external contour. The crystal partition function is
then defined by :%

\begin{equation}
\Xi^{\text{cr}}(\Gamma^{p})=\sum_{(X,Y)\in G(\Gamma^{p})}q^{-H_{V(\Gamma^{p}%
)}^{\text{dil}}(X)}\prod_{\gamma\in Y}\psi(\gamma) \label{eq:3.15}%
\end{equation}

\begin{lemma}
\label{L:I1} The following set of recurrence equations holds :
\begin{equation}
\Xi_{p}^{\text{dil}}(\Omega)=\sum_{\{\Gamma_{1}^{p},\ldots,\Gamma_{n}%
^{p}\}_{\text{ext}}}q^{-e_{p}\left\Vert \text{Ext}\right\Vert }\prod_{i=1}%
^{n}\Xi^{\text{cr}}(\Gamma_{i}^{p}) \label{eq:3.16}%
\end{equation}
Here the sum is over families $\{\Gamma_{1}^{p},\ldots,\Gamma_{n}%
^{p}\}_{\text{ext}}$ of mutually compatible external contours in
$\Omega$
($\text{supp}\,\Gamma_{i}^{p}\subset\Omega_{\text{int}}=\left\{
i\in \Omega:d(i,\mathbb{L}_{M}\setminus\Omega)>1\right\} $),
$\left\Vert \text{Ext}\right\Vert =\left\Vert \Omega\right\Vert
-\sum\limits_{i}\left\Vert V(\Gamma_{i}^{p})\right\Vert$ where
$\left\Vert V(\Gamma_{i}^{p})\right\Vert$ is the number of lines
of $V(\Gamma_{i}^{p})$;
\begin{equation}
\Xi^{\text{cr}}(\Gamma^{p})=\varrho(\Gamma^{p})\,\prod\limits_{m\in\left\{
\text{f},\text{fo}\right\}  }\Xi_{m}^{\text{dil}}(\text{Int}_{m}\Gamma^{p})\,
\label{eq:3.17}%
\end{equation}
where:
\begin{equation}
\varrho(\Gamma^{p})\,\equiv q^{-H_{\text{supp}\,\Gamma^{p}}^{\text{dil}%
}(X^{^{_{\Gamma^{p}}}})}\prod_{\gamma\in Y_{\Gamma^{p}}}\psi(\gamma)
\label{eq:3.18}%
\end{equation}

\end{lemma}

\begin{proof}
We have only to observe that for any $X\in\mathbf{H}^{p}(\Omega)$
\begin{equation}
H_{\Omega}^{\text{dil}}(X)=\sum_{\Gamma}H_{\text{supp}\,\Gamma}^{\text{dil}%
}(X^{\Gamma})+\sum_{p}e_{p}\left\Vert L_{p}(X)\cap\Omega\right\Vert
\end{equation}
where the sum is over all contours of the boundary of the configuration
$(X,Y=\emptyset)$ and $\left\Vert L_{p}(X)\cap\Omega\right\Vert $ is the
number of $p$--correct lines inside $\Omega$ of this configuration.
\end{proof}

Lemma \ref{L:I1} gives the following expansion for the partition function
\begin{equation}
\Xi_{p}^{\text{dil}}(\Omega)=q^{-e_{p}\left\Vert \Omega\right\Vert }%
\sum_{_{\{\Gamma_{1}^{p},\ldots,\Gamma_{n}^{p}\}_{_{\text{comp}}}}}\prod
_{i=1}^{n}z(\Gamma_{i}^{p}) \label{eq:3.20}%
\end{equation}
where the sum is now over families of compatibles contours of the same class
and
\begin{equation}
z(\Gamma^{p})=\varrho(\Gamma^{p})\,q^{e_{p}\left\Vert \Gamma^{p}\right\Vert
}\frac{\Xi_{m}^{\text{dil}}(\text{Int}_{m}\Gamma^{p})}{\Xi_{p}^{\text{dil}%
}(\text{Int}_{m}\Gamma^{p})} \label{eq:3.21}%
\end{equation}
with $m\neq p$ (and $\left\Vert \Gamma^{p}\right\Vert $ is the number of lines
of $\text{supp}\,\Gamma^{p}$).

To control the behavior of our system, we need to show Peierls
condition, that means that
$\varrho(\Gamma^{p})\,q^{e_{p}\left\Vert \Gamma^{p}\right\Vert }$
has good decaying properties with respect to the number of
incorrect points $\left\vert \Gamma^{p}\right\vert $ of
$\text{supp}\,\Gamma^{p}$. In fact we shall consider the modified
Peierls condition introduced in ref.\ \cite{KP2} where $e_p$ is
replaced by  $\underline{e}=\min\left(
e_{\text{f}},e_{\text{fo}}\right)$. Let
\begin{equation}
e^{-\tau}=\left(  2^{2d-1}q^{-\frac{1}{2(d-1)}}+24c\nu_{d}^{3}q^{-\left(
\frac{1}{d}-\beta_{b}\right)  }\right)  \frac{1}{1-6c\nu_{d}^{3}q^{-\left(
\frac{1}{d}-\beta_{b}\right)  }} \label{eq:3.23}%
\end{equation}
where $c=8e(e-1)(\sqrt{3}+2)e^{\frac{2}{\sqrt{3}+1}}$ and $\nu_{d}=(2d)^{2}$.
We have the following

\begin{proposition}
[Peierls condition]\label{P:PE1} Let $\Omega\subset\mathbb{L}_{M}$ be a finite
connected set of lines, assume that $\beta_{b}<\frac{1}{d}$, and $6c\nu
_{d}^{3}q^{-\frac{1}{d}+\beta_{b}}<1$, then for all $\beta_{s}\in\mathbb{R}$
:
\begin{equation}
\sum_{\Gamma^{p}:\text{supp}\,\Gamma^{p}=S}\left\vert \varrho(\Gamma
^{p})\,\right\vert \,q^{\underline{e}\left\Vert \Gamma^{p}\right\Vert }\leq
e^{-\tau\left\Vert S\right\Vert } \label{eq:3.24}%
\end{equation}
where $\left\Vert S\right\Vert $ is the number of lines of $S$.
\end{proposition}

The proof is postponed to the Appendix.

The recurrence equations of Lemma \ref{L:I1} play in our case the
role of Lemma 2.3 from Ref.\ \cite{S}. Together with the Peierls
estimates \ref{P:PE1}, they allow to study the states invariant
under horizontal translation (HTIS) of the hydra system by
applying Pirogov-Sinai theory. Here we do not need any extension
\cite{DS} \cite{BKL} developed for interacting contours. Actually
we have non interacting contours and only a standard form of the
theory is required. We present it in the next subsection. We chose
Zahradnik's approach \cite{Z1} with the notion of truncated
contour models including some improvements from \cite{KP}. We are
dealing with complex activities, a situation analyzed in Refs.\
\cite{Z2}, \cite{BI} and we mainly follow the presentation given
in that last reference.

\subsection{Diagram of horizontal translation invariant states}

To state our result, we first define the functional%
\begin{equation}
K_{p}(S)=\sum_{\Gamma^{p}:\text{supp}\,\Gamma^{p}=S}z(\Gamma^{p}%
)\label{eq:3.25}%
\end{equation}
Consider the partition function $\Xi_{p}^{\text{dil}}(\Omega)$ (\ref{eq:3.20})
and for a compatible family $\left\{  \Gamma_{1}^{p},...,\Gamma_{n}%
^{p}\right\}  _{\text{comp}}$ of $\ p$-contours, denote by $S_{1},...,S_{n}$
their respective supports. By summing over all contours with the same support
this partition function can be written as the partition function of a gas of
polymers $S$ with activity $K_{p}(S)=\sum\limits_{\Gamma^{p}:\text{supp}%
\,\Gamma^{p}=S}z(\Gamma^{p})$ interacting through hard-core exclusion
potential:%
\begin{equation}
\Xi_{p}^{\text{dil}}(\Omega)=q^{-e_{p}\left\Vert \Omega\right\Vert }%
\sum\limits_{\left\{  S_{1},...,S_{n}\right\}  _{\text{comp}}}\prod
\limits_{i=1}^{n}K_{p}(S_{i})\label{eq:3.26}%
\end{equation}
Here $\left\{  S_{1},...,S_{n}\right\}  _{\text{comp}}$ denotes compatible
families of polymers, that is \newline$d(S_{i},S_{j})>1$ for every two $S_{i}$
and $S_{j}$ in the family: recall that by definitions of contours a polymer
$S$ is a\ set of lines whose graph that joins all the points of the lines of
$S$ at distance $d(i,j)\leq1$ is connected.

Next we define auxiliary models, namely the truncated contour
models associated to each boundary conditions.

\begin{definition}
\label{D:TCF}A truncated contour functional is defined as%
\begin{equation}
K_{p}^{\prime}(S)=\left\{
\begin{array}
[c]{ll}%
K_{p}(S) & \text{if }\left\Vert K_{p}(S)\right\Vert \leq e^{-\alpha\left\Vert
S\right\Vert }\text{ }\\
0 & \text{otherwise}%
\end{array}
\right.  \label{eq:3.27}%
\end{equation}
where $\left\Vert K_{p}(S)\right\Vert =\sum_{\Gamma^{p}:\text{supp}%
\,\Gamma^{p}=S}\left\vert z(\Gamma^{p})\right\vert $ , and
$\alpha>0$ is some positive parameter to be chosen later (see
Theorem~\ref{T:unicity} below).
\end{definition}

\begin{definition}
\label{D:stability1}The collection $\left\{  S,p\right\}  $ of all
$p$-contours $\Gamma^{p}$with support \newline supp$\,\Gamma^{p}=S$ is called
stable if%
\begin{equation}
\left\Vert K_{p}(S)\right\Vert \leq e^{-\alpha\left\Vert S\right\Vert }
\label{eq:3.28}%
\end{equation}
i.e. if $K_{p}(S)=K_{p}^{\prime}(S)$.
\end{definition}

We define the truncated partition function $\Xi_{p}^{\prime}(\Omega)$ as the
partition function obtained from $\Xi_{p}^{\text{dil}}(\Omega)$ by leaving out
unstable collections of contours, namely%

\begin{align}
\Xi_{p}^{\prime}(\Omega) &  =q^{-e_{p}\left\Vert \Omega\right\Vert
}\sideset{}{'}\sum_{\{\Gamma_{1}^{p},\ldots,\Gamma_{n}^{p}\}_{\text{comp}}%
}\prod_{i=1}^{n}z(\Gamma_{i}^{p})\nonumber\\
&  =q^{-e_{p}\left\Vert \Omega\right\Vert }\sum\limits_{\left\{
S_{1},...,S_{n}\right\}  _{\text{comp}}}\prod\limits_{i=1}^{n}K_{p}^{\prime
}(S_{i})\label{eq:3.29}%
\end{align}
Here the sum goes over compatible families of \textit{stable collections of
contours}. Let
\begin{equation}
h_{p}=-\lim_{\Omega\rightarrow L}\frac{1}{\left\Vert \Omega\right\Vert }\ln
\Xi_{p}^{\prime}(\Omega)\label{eq:3.31}%
\end{equation}
be the \textit{\ metastable free energy } of the truncated partition function
$\Xi_{p}^{\prime}(\Omega)$.

For $\alpha$ large enough, the thermodynamic limit
(\ref{eq:3.31}) can be controlled by a convergent cluster
expansion. We conclude the existence of $h_{p}$, together
with the bounds%
\begin{align}
e^{-\kappa e^{-\alpha}\left\vert \partial_{s}\Omega\right\vert } &  \leq
\Xi_{p}^{\prime}(\Omega)e^{h_{p}\left\Vert \Omega\right\Vert }\leq e^{\kappa
e^{-\alpha}\left\vert \partial_{s}\Omega\right\vert }\label{eq:3.33}\\
\left\vert h_{p}-e_{p}\ln q\right\vert  &  \leq\kappa e^{-\alpha
}\label{eq:3.34}%
\end{align}
shown below. Here $\kappa=\kappa_{\text{cl }}(\chi^{\prime})^{2}$
where $\kappa_{\text{cl
}}=\frac{\sqrt{5}+3}{2}e^{\frac{2}{\sqrt{5}+1}}$ is the cluster
constant \cite{KP} and $\kappa^{\prime}=3^{d-1}-1$;
$\partial_{s}\Omega=\partial \Omega\cap\mathbb{L}_{0}$ in the way
defined in Subsection \ref{S:1.1}.

\begin{theorem}
\label{T:unicity} Assume $\beta_{b}<1/d$ and $q$ is large enough so that
$e^{-\alpha}\equiv\newline e^{-\tau+2\kappa^{\prime}+3} <\frac{0.7}%
{\kappa\kappa^{\prime}}$, then there exists a unique $\beta_{s}^{t}=\frac
{1}{d-1}+O(e^{-\tau})$ such that :

\begin{description}
\item[(i)] for $\beta_{s}=\beta_{s}^{t}$%
\[
\Xi_{p}^{\text{dil}}(\Omega)=\Xi_{p}^{\prime}(\Omega)
\]
for both boundary conditions $p=$f and $p=$of , and the free energy of the
hydra model is given by $g_{\text{f}}+\ln q=h_{\text{f }}=h_{\text{fo }}$ .

\item[(ii)] for$\ \beta_{s}<\beta_{s}^{t}$
\[
\Xi_{\text{f}}^{\text{dil}}(\Omega)=\Xi_{\text{f}}^{\prime}(\Omega)
\]
and $g_{\text{f}}+\ln q=h_{\text{f }}<h_{\text{fo }}$

\item[(iii)] for$\ \beta_{s}>\beta_{s}^{t}$%
\[
\Xi_{\text{fo}}^{\text{dil}}(\Omega)=\Xi_{\text{fo}}^{\prime}(\Omega)
\]
and $g_{\text{f}}+\ln q=h_{\text{fo }}<h_{\text{f }}$
\end{description}
\end{theorem}

Let us introduce the Gibbs states $\langle\,\cdot\,\rangle^{p}$, associated to
the partition functions $\Xi_{p}^{\text{dil}}$. Theorems above show also that
at $\beta_{s}=\beta_{s}^{t}$%
\begin{equation}%
\begin{array}
[c]{ll}%
\langle\,n_{b}\,\rangle^{\text{f}} & =O(e^{-\tau})\\
\langle\,n_{b}\,\rangle^{\text{fo}} & =1-O(e^{-\tau})
\end{array}
\label{eq:3.41}%
\end{equation}
for any bond $b$ of the boundary layer while
\begin{equation}%
\begin{array}
[c]{ll}%
\langle\,n_{b^{\prime}}\,\rangle^{\text{f}} & =O(e^{-\tau})\\
\langle\,n_{b^{\prime}}\,\rangle^{\text{fo}} & =O(e^{-\tau})
\end{array}
\label{eq:3.42}%
\end{equation}
for any bond $b^{\prime}$ \ between the boundary layer and the
first layer. This is because, with free ($\text{f}$) boundary
conditions, bonds $b$ of the boundary layer and bonds $b^{\prime}$
of between the boundary and first layer are occupied only if there
is a free contour surrounding them and that the correlation
functions are controlled  by the contour model cluster expansion.
With free-ordered (fo) boundary conditions, bonds $b$ are empty
and bonds $b^{\prime}$ are occupied only if there is a
$\text{fo}$-contour surrounding it and again the correlations are
controlled by cluster expansion.

This shows in particular that the derivative $\frac{\partial}{\partial
K}g_{\text{f}}$ of the free energy $g_{\text{f}}$ with respect to the surface
coupling constant $K$ is discontinuous near $K=\beta^{-1}\ln\left(
1+q^{\frac{1}{d-1}}\right)  $.

Obviously, the first relations of (\ref{eq:3.41}) and (\ref{eq:3.42}) hold
true for any $\beta_{s}\leq\beta_{s}^{t}$ while the second relations of
(\ref{eq:3.41}) and (\ref{eq:3.42}) hold true for any $\beta_{s}\geq\beta
_{s}^{t}$. Notice also that the estimate (\ref{eq:3.42}) on the mean values of
the bonds $b^{\prime}$ between the boundary and first layer can be extended to
the bonds of the bulk.

\textbf{Outline of proof of Theorem~\ref{T:unicity}}

\bigskip The first step is to prove the inequalities (\ref{eq:3.33}) and
(\ref{eq:3.34}). To exponentiate the partition function
\begin{equation}
\mathcal{Z}_{p}(\Omega)=q^{e_{p}\left\Vert \Omega\right\Vert }\Xi_{p}^{\prime
}(\Omega)=\sum\limits_{\left\{  S_{1},...,S_{n}\right\}  _{\text{comp}}}%
\prod\limits_{i=1}^{n}K_{p}^{\prime}(S_{i})
\end{equation}
we define the truncated functional $\Phi^{T}$ associated to $K_{p}^{\prime}$
\begin{equation}
\Phi^{T}(X)=\frac{a(X)}{\prod_{\gamma}X(S)!}\prod_{S}K_{p}^{\prime}(S)^{X(S)}%
\end{equation}
defined on the multi-indexes $X$ associated to the polymers: a multi-index is
a function from the set of polymers into the set of non negative integers, and
$a(X)$ is defined as in (\ref{eq:2.11}).

The number of polymers $S$ with number of lines $\left\Vert S\right\Vert =n$
and containing a given line can be bounded by $\nu^{n}$ where $\nu
=(3^{d-1}-1)^{2}$. Actually, There exists at most $3^{d-1}-1$ lines at
distance $1$ of a given line and for any connected set of lines there exists a
graph going through every line at most twice.

As a result of the standard cluster expansion \cite{KP,M}, we get for $\kappa
e^{-\alpha}<1$
\begin{align}
\mathcal{Z}_{p}(\Omega)  &  =\exp\left\{  \sum\limits_{X}\Phi_{T}(X)\right\}
\label{eq:B5}\\
&  =\exp\left\{  \left\Vert \Omega\right\Vert \sum\limits_{X:\text{supp}%
X\backepsilon L}\frac{\Phi^{T}(X)}{\left\Vert \text{supp}X\right\Vert
}
-\sideset{}{^{*}}\sum_{X}
\frac{\left\Vert \text{supp}X\cap \Omega\right\Vert}{
\left\Vert \text{supp}X\right\Vert
}
\Phi^{T}(X)
\right\}
\label{eq:B6}%
\end{align}
Here, the sum in (\ref{eq:B5}) is over multi-indexes whose support$\ $
$\text{supp }X=\left\{  S:X(S)\newline\geq1\right\}  $ belongs to $\Omega$,
the first sum in (\ref{eq:B6}) is over all multi-indexes $X$ whose support
contains a given line $L$, and the second sum in (\ref{eq:B6}) runs over
multi-indexes whose support intersects both $\Omega$ and its complement
$\mathbb{L}_{M}\setminus\Omega$. \newline The series $\sum
\limits_{X:\text{supp}X\backepsilon L}\left\vert \Phi^{T}(X)\right\vert $ is
absolutely convergent and satisfies the bound
\begin{equation}
\sum\limits_{X:\text{supp}X\backepsilon L}\left\vert \Phi^{T}(X)\right\vert
\leq\kappa e^{-\alpha}%
\end{equation}
that immediately gives (\ref{eq:3.33}) and (\ref{eq:3.34}).

The second step closely follows the Borgs Imbrie paper \cite{BI}.

We put
\begin{equation}
a_{p}=\text{ }h_{p}-\min_{m}\text{ }h_{m} \label{eq:3.32}%
\end{equation}
The boundary condition $p$ is then called stable if $a_{p}=0$. Our task is now
to show that if the boundary condition $p$ is stable, then all collections of
$p$--contours are stable implying that $\Xi_{p}^{\prime}(\Omega)$ coincides
with $\Xi_{p}^{\text{dil}}(\Omega)$.

We notice that when $a_{p}\leq1$ and the hypotheses of Proposition \ref{P:PE1}
hold, then the condition%
\begin{equation}
\frac{\Xi_{m}^{\text{dil}}(\text{Int}_{m}\Gamma^{p})}{\Xi_{p}^{\text{dil}%
}(\text{Int}_{m}\Gamma^{p})}\leq e^{2\left\vert \partial_{s}\text{Int}%
_{m}\Gamma^{p}\right\vert } \label{eq:3.35}%
\end{equation}
for all contours $\Gamma^{p}$ with support $S$ ensures that the collection
$\left\{  S,p\right\}  $ is stable, provided
\begin{equation}
e^{-\alpha}= e^{-\tau+2\kappa^{\prime}+3}<\frac{1}{\kappa} \label{eq:3.36}%
\end{equation}
Indeed since%
\begin{equation}
z(\Gamma^{p})=\varrho(\Gamma^{p})q^{\underline{e}\left\Vert \Gamma
^{p}\right\Vert }q^{\left(  e_{p}-\underline{e}\right)  \left\Vert \Gamma
^{p}\right\Vert }\frac{\Xi_{m}^{\text{dil}}(\text{Int}_{m}\Gamma^{p})}{\Xi
_{p}^{\text{dil}}(\text{Int}_{m}\Gamma^{p})} \label{eq:3.37}%
\end{equation}
we get, by Proposition \ref{P:PE1}, using ($e_{p}-\underline{e})\ln q\leq
a_{p}+2\kappa e^{-\alpha}\leq3$ (see (\ref{eq:3.34})), and the inequality
$\left\vert \partial_{s}\text{Int}_{m}\Gamma^{p}\right\vert \leq\kappa
^{\prime}\left\Vert \Gamma^{p}\right\Vert $, that $\left\Vert K_{p}%
(S)\right\Vert \leq e^{-\tau\left\Vert S\right\Vert }e^{(3+2\kappa^{\prime
})\left\Vert S\right\Vert }$.

We then show inductively on $\text{diam}\, \Omega=\max_{i,j\in\Omega
\cap\mathbb{L}_{0}}d(i,j)$, that

\begin{description}
\item[(i)] if $a_{m}$diam$\,\Omega\leq1$, and $a_{p}=0$, then
\begin{equation}
\left\vert \frac{\Xi_{P}^{\text{dil}}(\Omega)}{\Xi_{m}^{\text{dil}}(\Omega
)}\right\vert \leq e^{a_{m}\left\Vert \Omega\right\Vert +2\kappa e^{-\alpha
}\left\vert \partial_{s}\Omega\right\vert } \label{eq:C1}%
\end{equation}

\item[(ii)] if $a_{p}=0$, then%
\begin{equation}
\left\vert \frac{\Xi_{m}^{\text{dil}}(\Omega)}{\Xi_{p}^{\text{dil}}(\Omega
)}\right\vert \leq e^{3\kappa e^{-\alpha}\left\vert \partial_{s}%
\Omega\right\vert } \label{eq:C2}%
\end{equation}

\item[(iii)] if $a_{m} \text{diam}\,\Omega\leq1$, then
\begin{equation}
\left\vert \frac{\Xi_{\widetilde{m}}^{\text{dil}}(\Omega)}{\Xi_{m}%
^{\text{dil}}(\Omega)}\right\vert \leq e^{(1+5\kappa e^{-\alpha})\left\vert
\partial_{s}\Omega\right\vert }%
\end{equation}

\end{description}

The proof is analog to that of Theorem 3.1 in \cite{BI} using our previous
estimates and thus we omit it. It gives that for $\beta_{b}<1/d$ and $q$ is
large enough so that $e^{-\tau+2\kappa^{\prime}+3}=e^{-\alpha}<\frac
{1}{5\kappa}$, then if $a_{p}=0$ all collections of $p$-contours are stable.

For our last step, it remains to show that there exists a unique $\beta
_{s}^{t}=\frac{1}{d-1}+O(e^{-\tau})$ such that:%
\begin{align}
a_{\text{f}}  &  =0\ \text{and}\ a_{\text{fo}}=0\hspace*{2cm}\text{for}%
\ \beta_{s}=\beta_{s}^{t}\label{eq:R1}\\
a_{\text{f}}  &  =0\ \text{and}\ a_{\text{fo}}>0\hspace*{2cm}\text{for}%
\ \beta_{s}<\beta_{s}^{t}\label{eq:R2}\\
a_{\text{f}}  &  >0\ \text{and}\ a_{\text{fo}}=0\hspace*{2cm}\text{for}%
\ \beta_{s}>\beta_{s}^{t} \label{eq:R3}%
\end{align}
This is a consequence of the fact that the free energy \ $\frac{-1}{\left\Vert
\Omega\right\Vert \ln q}\ln\mathcal{Z}_{p}(\Omega)$ is a Lipschitz function of
$\beta_{s}$ uniformly in $\Omega$\ with a small Lipschitz constant (when
$\tau$ is large, one has a good control on the one sided derivative
$\frac{\partial}{\partial^{\pm}\beta_{s}}K_{p}^{\prime}(S)$ of the activities
associated to the partition functions $\mathcal{Z}_{p}(\Omega)$).

\begin{lemma}
\label{L:Lif} Assume that $\kappa e^{-\alpha}<\frac{c}{\kappa^{\prime}}%
\leq\frac{0.7}{\kappa^{\prime}}$, then%
\begin{equation}
\left\vert \frac{1}{\left\Vert \Omega\right\Vert \ln q}\frac{\partial
}{\partial^{\pm}\beta_{s}}\ln\mathcal{Z}_{p}(\Omega)\right\vert \leq
(d-1)\frac{a_{c}\kappa e^{-\alpha}}{1-a_{c}\kappa e^{-\alpha}} \label{eq:D1}%
\end{equation}
where $a_{c}=\frac{e}{\kappa_{\text{cl }}}e^{9c/8}$ .
\end{lemma}

\begin{proof}

By virtue of relation (\ref{eq:3.21}) one gets
\begin{align}
\frac{\partial}{\partial\beta_{s}}K_{p}(S)=\sum\limits_{\Gamma^{p}%
:\text{supp}\,\Gamma^{p}=S}  &  \Bigg(\frac{\partial}{\partial\beta_{s}}%
\ln\varrho(\Gamma^{p})\,+\frac{\partial}{\partial\beta_{s}}\ln q^{e_{p}%
\left\Vert \Gamma^{p}\right\Vert }\nonumber\\
&  \hphantom{xxxxxxx}+\frac{\partial}{\partial\beta_{s}}\ln\frac{\Xi
_{m}^{\text{dil}}(\text{Int}_{m}\Gamma^{p})}{\Xi_{p}^{\text{dil}}%
(\text{Int}_{m}\Gamma^{p})}\Bigg)z(\Gamma^{p}) \label{eq:D3}%
\end{align}
Since $\frac{\partial e_{p}}{\partial\beta_{s}}=0$ or $\frac{\partial e_{p}%
}{\partial\beta_{s}}=-(d-1)$ (see (\ref{eq:3.12})), the bound (\ref{eq:NA.15})
on the first term inside the parenthesis gives that $ \left\vert
\frac{\partial}{\partial\beta_{s}}\ln\varrho(\Gamma^{p})+\frac{\partial
}{\partial\beta_{s}}\ln q^{e_{p}\left\Vert \Gamma^{p}\right\Vert }\right\vert
\leq\left\Vert \Gamma^{p}\right\Vert (d-1)\ln q $. This inequality together
with the bound (\ref{eq:3.14}) on the third term inside the parenthesis leads
to%
\begin{align}
\left\vert \frac{\partial}{\partial\beta_{s}}K_{p}(S)\right\vert  &  \leq
\sum\limits_{\Gamma^{p}:\text{supp}\,\Gamma^{p}=S}[\left\Vert \Gamma
^{p}\right\Vert (d-1)\ln q+\left\Vert \text{Int}_{m}\text{ }\Gamma
^{p}\right\Vert (d-1)\ln q]\left\vert z(\Gamma^{p})\right\vert \nonumber\\
&  \leq\left[  \left\Vert V(S)\right\Vert (d-1)\ln q\right]  e^{-\alpha
\left\Vert S\right\Vert } \label{eq:D4}%
\end{align}
where $V(S)\equiv V(\Gamma^{p})=$Int $\Gamma^{p}\cup\Gamma^{p}$. \newline
Taking into account the fact that either the one-sided derivative
$\frac{\partial}{\partial^{\pm}\beta_{s}}K_{p}^{\prime}(S)=0$ or
$\frac{\partial}{\partial^{\pm}\beta_{s}}K_{p}^{\prime}(S)\leq\frac{\partial
}{\partial\beta_{s}}K_{p}(S)$ we get that the left-hand side of (\ref{eq:D1})
is bounded as
\begin{align}
\bigg\vert\frac{1}{\left\Vert \Omega\right\Vert }  &  \frac{\partial}%
{\partial^{\pm}\beta_{s}}\ln\mathcal{Z}_{p}(\Omega)\bigg\vert\leq
\sum\limits_{S:S\backepsilon L}\left\vert \frac{\partial}{\partial^{\pm}%
\beta_{s}}K_{p}^{\prime}(S)\right\vert \times\left\vert \frac
{\sideset{}{^{*}}\sum_{\left\{  S_{1},...,S_{n}\right\}  _{\text{comp}}}%
K_{p}^{\prime}(S)}{\mathcal{Z}_{p}(\Omega)}\right\vert \nonumber\\
&  \leq(d-1)\ln q\sum\limits_{S:S\backepsilon L}\left\Vert V(S)\right\Vert
e^{-\alpha\left\Vert S\right\Vert }\exp\left\{  -\sum\limits_{X:\text{supp}%
X\cap\overline{S}=\varnothing}\Phi^{T}(X)\right\} \nonumber\\
&  \leq(d-1)\ln q\sum\limits_{S:S\backepsilon L}\left\Vert V(S)\right\Vert
e^{-\alpha\left\Vert S\right\Vert }e^{(1+\kappa^{\prime})\kappa e^{-\alpha
}\left\Vert S\right\Vert } \label{eq:D5}%
\end{align}
where the sum $\sideset{}{^{*}}\sum$ is over all families $\left\{
S_{1},...,S_{n}\right\}  _{\text{comp}}$ compatible with $S$ and $\overline
{S}=S\cup\left\{  L:d(L,S)=1\right\}  $ is the set of lines at distance less
or equal to $1$ from $S$.

The last inequality leads to the lemma by inserting the bound%
\[
\left\Vert V(S)\right\Vert \leq\left\Vert S\right\Vert \text{diam }%
S\leq\left\Vert S\right\Vert ^{2}\leq e^{\left\Vert S\right\Vert }
\]
and using the fact that the number of polymers with number of lines
$\left\Vert S\right\Vert =n$ and containing a given line can be bounded by
$\nu^{n}=(\kappa^{\prime})^{2n}$.
\end{proof}

To prove relations (\ref{eq:R1}-\ref{eq:R3}), we consider the energy
difference $\frac{e_{\text{fo}}-e_{\text{f}}}{d-1}=\beta_{s}-\frac{1}{d-1}$.
As a function of $\beta_{s}$ it is obviously increasing, negative for
$\beta_{s}<\frac{1}{d-1}$, positive for $\beta_{s}>\frac{1}{d-1}$, moreover it
intersects the horizontal coordinate axis only at one point $\beta_{s}%
=\frac{1}{d-1}$. \newline Since $\frac{a_{\text{fo}}-a_{\text{f}}}{\left(
d-1\right)  \ln q}$ differs only from $\frac{e_{\text{fo}}-e_{\text{f}}}{d-1}$
by a Lipschitz function of $\beta_{s}$, namely the function $\lim
_{\Omega\uparrow\mathbb{L}}\frac{1}{\left\Vert \Omega\right\Vert (d-1)\ln
q}\left[  \ln\mathcal{Z}_{\text{f}}(\Omega)-\ln\mathcal{Z}_{\text{fo}}%
(\Omega)\right]  $, it will satisfy the same properties (the intersecting
point slightly changed) provided the Lipschitz constant of this last function
is sufficiently small. Indeed,
\begin{equation}
\frac{a_{\text{fo}}-a_{\text{f}}}{(d-1)\ln q}=\frac{e_{\text{fo}}-e_{\text{f}%
}}{d-1}-\lim_{\Omega\uparrow\mathbb{L}}\frac{1}{\left\Vert \Omega\right\Vert
(d-1)\ln q}\left[  \ln\mathcal{Z}_{\text{fo}}(\Omega)-\ln\mathcal{Z}%
_{\text{f}}(\Omega)\right]  \label{eq:D2}%
\end{equation}
By virtue of Lemma \ref{L:Lif}, the needed condition on the Lipschitz constant
is $2\frac{a_{c}\kappa e^{-\alpha}}{1-a_{c}\kappa e^{-\alpha}}<1$ since the
Lipschitz constant of $\frac{e_{\text{fo}}-e_{\text{f}}}{d-1}$ is 1. This last
inequality requires $a_{c}\kappa e^{-\alpha}<1/3$ and is actually fulfilled
under the hypotheses of the Theorem \ref{T:unicity}: by noticing that the
geometric constant $\kappa^{\prime}=3^{d-1}-1\geq8$, these hypotheses imply
$a_{c}\kappa e^{-\alpha}\leq a_{c}c/\kappa^{\prime}\leq a_{c}c/8$ where the
last term is less than $1/3$ for $c\leq0.7$. {\rule{0.5em}{0.5em}}

\subsection*{\bigskip\ Acknowledgments}

The authors thank S. Miracle-Sol\'{e}, S. Shlosman, and V. Zagrebnov for
helpful discussions. L.L. acknowledges the\ warm hospitality and financial
support of the Centre de Physique Th\'{e}orique.

\newpage

\section*{ Appendix: Proof of Proposition \ref{P:PE1}}

\renewcommand{\theequation}{A.\arabic{equation}} \renewcommand{\thesection}{A}
\setcounter{equation}{0} \setcounter{theorem}{0}

We begin the proof by considering contours $\Gamma=\left\{  \text{supp}%
\,\Gamma,X^{\Gamma},Y^{\Gamma}\right\}  $ (where $\left\{  X^{\Gamma
},Y^{\Gamma}\right\}  $ is the configuration having $\Gamma$ as unique
contour) without polymers, i.e. $Y^{\Gamma}=\emptyset$. We let $X_{s}^{\Gamma
}=X^{\Gamma}\cap B(\mathbb{L}_{0})$ be the bonds of $X^{\Gamma} $\ that belong
to the boundary layer and $X_{b}^{\Gamma}=X^{\Gamma}\setminus X_{s}^{\Gamma}$.
A site $i\in\mathbb{L}_{0}$ of the boundary layer will be called regular if
the bonds of its envelope that belong to the boundary layer are either all
empty or all occupied. It will be called irregular otherwise and we will
denote by $I_{0}(\Gamma)$ the set of irregular sites of the contour $\Gamma$.
Namely $I_{0}(\Gamma)=\left\{  i\in\mathbb{L}_{0}:1\leq\left\vert E(i)\cap
X_{s}^{\Gamma}\right\vert \leq2(d-1)-1\right\}  $.

\begin{lemma}
\label{AP1}%
\begin{equation}
\left\vert \varrho(\Gamma^{p})\,\right\vert \,q^{\underline{e}\left\Vert
\Gamma^{p}\right\Vert }\leq q^{-\frac{\left\vert I_{0}(\Gamma^{p})\right\vert
}{2(d-1)}-\left(  \frac{1}{d}-\beta_{b}\right)  \left\vert X_{b}^{\Gamma^{p}%
}\right\vert } \label{eq:A.1}%
\end{equation}

\end{lemma}

\begin{proof}
By Lemma \ref{L:I1} and definition (\ref{eq:3.9}), one has%
\begin{align}
\varrho(\Gamma^{p})\,q^{\underline{e}\left\Vert \Gamma^{p}\right\Vert }  &
=q^{-(H_{\text{supp}\,\Gamma^{p}}^{\text{dil}}(X)-\underline{e}\left\Vert
\Gamma^{p}\right\Vert )}\nonumber\\
&  =q^{\sum_{i\in\text{supp}\,\Gamma^{p}\cap
B(\mathbb{L}_{0})}\frac{\beta _{s}}{2}\left\vert E(i)\cap
X_{s}\right\vert +\beta_{b}\left\vert X_{b}\right\vert
-S(X)+C(X)-\chi_{p}+\underline{e}\left\Vert \Gamma^{p}\right\Vert
} \label{eq:A.2}%
\end{align}
Here, $\chi_{\text{f}}=0$, $\chi_{\text{o}}=1$, and to simplify formulae we
put hereafter $X,X_{s}$, and $X_{b}$ instead of $X^{\Gamma^{p}},X_{s}%
^{\Gamma^{p}}$, and $X_{b}^{\Gamma^{p}}$. By using the relations%
\begin{align}
S(X)  &  =S(X_{s})+S(X_{b}\cup X_{s})-S(X_{s})\label{eq:A.3}\\
C(X)  &  =C(X_{s})+C(X_{b}\cup X_{s})-C(X_{s}) \label{eq:A.4}%
\end{align}
we get%
\begin{equation}
\varrho(\Gamma^{p})\,q^{\underline{e}\left\Vert \Gamma^{p}\right\Vert }\leq
q^{\mathbf{A}_{s}(\Gamma^{p})+\mathbf{A}_{b}(\Gamma^{p})} \label{eq:A.5}%
\end{equation}
where%
\begin{align}
\mathbf{A}_{s}(\Gamma^{p})  &  =\sum_{i\in\text{supp}\,\Gamma^{p}\cap
B(\mathbb{L}_{0})}\left(  \frac{\beta_{s}}{2}\left\vert E(i)\cap
X_{s}\right\vert -\left\vert i\cap S(X)\right\vert +\underline{e}\right)
+C(X_{s})\label{eq:A.6}\\
\mathbf{A}_{b}(\Gamma^{p})  &  =\beta_{b}\left\vert X_{b}\right\vert
-S(X_{b}\cup X_{s})+S(X_{s})+C(X_{b}\cup X_{s})-C(X_{s}) \label{eq:A.7}%
\end{align}
Let us first bound $\mathbf{A}_{s}(\Gamma^{p})$.

Consider an irregular site (of the boundary layer). Then by definition:
$1\leq\left\vert E(i)\cap X_{s}\right\vert \leq2(d-1)-1$. Since \underline{$e
$}$=\min\left\{  e_{\text{f}}=0,e_{\text{fo}}=1-(d-1)\beta_{s}\right\}  $, we
have,%
\begin{align}
\frac{\beta_{s}}{2}\left\vert E(i)\cap X_{s}\right\vert -\left\vert i\cap
S(X)\right\vert +\underline{e}  &  =\frac{\beta_{s}}{2}\left\vert E(i)\cap
X_{s}\right\vert -1+\underline{e}\nonumber\\
&  \leq\frac{1-\underline{e}}{2(d-1)}\left\vert E(i)\cap X_{s}\right\vert
-1+\underline{e}\nonumber\\
&  =(1-\underline{e})\frac{\left\vert E(i)\cap X_{s}\right\vert -2(d-1)}%
{2(d-1)}\nonumber\\
&  \leq\frac{\left\vert E(i)\cap X_{s}\right\vert -2(d-1)}{2(d-1)}
\label{eq:A.8}%
\end{align}
where we used successively $\beta_{s}\leq(1-e)/(d-1)$ and $\underline{e}\leq0
$

Next, we take into account that the number of connected components may be
bounded as (c.f.\ e.g.\ \cite{KLMR,LMR}):
\begin{equation}
C(X)\leq\sum_{i:1\leq\left\vert E(i)\cap X_{s}\right\vert \leq d-1}\frac
{1}{2^{\left\vert E(i)\cap X_{s}\right\vert }} \label{eq:A.9}%
\end{equation}
When $1\leq\left\vert E(i)\cap X_{s}\right\vert \leq2(d-1)-1$, we infer
\begin{equation}
\frac{\left\vert E(i)\cap X_{s}\right\vert -2(d-1)}{2(d-1)}+\frac{\chi
(1\leq\left\vert E(i)\cap X_{s}\right\vert \leq d-1)}{2^{\left\vert E(i)\cap
X_{s}\right\vert }}\leq\frac{-1}{2(d-1)} \label{eq:A.10}%
\end{equation}
so that each irregular site $i$ of the boundary layer provides at most a
contribution $\frac{-1}{2(d-1)}$ to the right-hand side of (\ref{eq:A.6})
giving%
\begin{equation}
q^{\mathbf{A}_{s}(\Gamma^{p})}\leq q^{-\frac{\left\vert I_{0}(\Gamma
^{p})\right\vert }{2(d-1)}} \label{eq:A.11}%
\end{equation}
Consider now, the quantity $\mathbf{A}_{b}(\Gamma^{p})$. One has%
\begin{equation}
q^{\mathbf{A}_{b}(\Gamma^{p})}\leq\prod_{\delta}q^{\beta_{b}\left\vert
\delta\right\vert -S(\delta\cup X_{s})+S(X_{s})+C(\delta\cup X_{s})-C(X_{s})}
\label{eq:A.12}%
\end{equation}
where the product runs over connected component of $X_{b}$. By the same
geometrical observation as in the proof of Theorem~\ref{CE1}, one easily gets
the inequality%
\begin{equation}
\frac{\left\vert \delta\right\vert }{d}-S(\delta\cup X_{s})+S(X_{s}%
)+C(\delta\cup X_{s})-C(X_{s})\leq0 \label{eq:A.13}%
\end{equation}
that gives%
\begin{equation}
q^{\mathbf{A}_{b}(\Gamma^{p})}\leq q^{-(\frac{1}{d}-\beta_{b})\left\vert
X_{b}\right\vert } \label{eq:A.14}%
\end{equation}
which in turn implies the lemma by taking into account (\ref{eq:A.5})
and
(\ref{eq:A.11}).

Notice that by (\ref{eq:A.2}) one has the following bound%
\begin{equation}
\frac{\partial}{\partial\beta_{s}}\ln\varrho(\Gamma^{p})\,\leq\left\Vert
\Gamma^{p}\right\Vert (d-1)\ln q \label{eq:NA.15}%
\end{equation}

\end{proof}

Considering still contours $\Gamma=\left\{  \sup\Gamma,X^{\Gamma},Y^{\Gamma
}\right\}  $ without polymers, i.e. ($Y^{\Gamma}=\emptyset$) we have the

\begin{lemma}
\label{AP2}Assume that $\beta_{b}<\frac{1}{d}$, and $q^{\frac{1}{d}-\beta_{b}%
}>2\nu_{d}=2(2d)^{2}$, then
\begin{equation}
\sum_{\Gamma^{p}:\text{supp}\,\Gamma^{p}=S}\varrho(\Gamma^{p}%
)\,\,q^{\underline{e}\left\Vert \Gamma^{p}\right\Vert }\leq\left(
2^{2d-1}q^{-\frac{1}{2(d-1)}}+8\nu_{d}q^{-\left(  \frac{1}{d}-\beta_{b}\right)
}\right)  ^{\left\Vert S\right\Vert }\frac{1}{1-2\nu_{d}q^{-\left(  \frac
{1}{d}-\beta_{b}\right)  }} \label{eq:A.15}%
\end{equation}

\end{lemma}

which shows that, whenever $q$ is large enough, the Peierls condition holds
true for the class of contours without polymers.

\begin{proof}
First, observe that for contours $\Gamma^{p}$ with support supp$\,\Gamma
^{p}=S$ and number of irregular sites of the boundary layer $\left\vert
I_{0}(\Gamma^{p})\right\vert =k$ one has $\left\vert X_{b}\right\vert
=\left\vert \delta_{1}\right\vert +...+\left\vert \delta_{m}\right\vert
\geq\left\Vert S\right\Vert -k$. Therefore,
\begin{align}
\sum_{\Gamma^{p}:\text{supp}\,\Gamma^{p}=S}\varrho(\Gamma^{p})\,q^{\underline
{e}\left\Vert \Gamma^{p}\right\Vert }  &  \leq\sum_{0\leq k\leq\left\Vert
S\right\Vert }\sum_{\Gamma^{p}:\left\vert I_{0}(\Gamma^{p})\right\vert
=k}q^{-\frac{k}{2(d-1)}}q^{-(\frac{1}{d}-\beta_{b})\left\vert X_{b}\right\vert
}\nonumber\\
&  \leq\sum_{0\leq k\leq\left\Vert S\right\Vert }\binom{\left\Vert
S\right\Vert }{k}2^{2(d-1)k}2^{\left\Vert S\right\Vert -k}q^{-\frac{k}{2(d-1)}%
}\nonumber\\
&  \hphantom{xxx}\times\sum_{n\leq\left\Vert S\right\Vert }\sum
_{\substack{i_{1},...,i_{n};\\i_{\alpha}\in S\cap\mathbb{L}_{0}}%
}\sum_{\substack{\delta_{1}\backepsilon i_{1},...,\delta_{n}\backepsilon
i_{n}\\\left\vert \delta_{1}\right\vert +...+\left\vert \delta_{n}\right\vert
\geq\left\Vert S\right\Vert -k}}\prod_{j=1}^{m}q^{-\left(  \frac{1}{d}%
-\beta_{b}\right)  \left\vert \delta_{j}\right\vert } \label{eq:A.16}%
\end{align}
Here the binomial coefficient $\binom{\left\Vert S\right\Vert }{k}$ bounds the
choice of irregular sites of the boundary layer while the factor
$2^{2(d-1)k}2^{\left\Vert S\right\Vert -k}$ bounds the numbers of contours with
$\left\Vert S\right\Vert $ lines and $k$ irregular sites. Then%

\begin{align}
&  \sum_{\Gamma^{p}:\text{supp}\,\Gamma^{p}=S}\varrho(\Gamma^{p}%
)\,q^{\underline{e}\left\Vert \Gamma^{p}\right\Vert }\leq\sum_{0\leq
k\leq\left\Vert S\right\Vert }\binom{\left\Vert S\right\Vert }{k}%
2^{2(d-1)k}2^{\left\Vert S\right\Vert -k}q^{-\frac{k}{2(d-1)}}\nonumber\\
&  \hphantom{xxxx}\times\sum_{n\leq\left\Vert S\right\Vert }\binom{\left\Vert
S\right\Vert }{n}\sum_{m_{1}+...+m_{n}\geq\left\Vert S\right\Vert -k}%
\prod_{j=1}^{n}\left(  \nu_{d}q^{-\left(  \frac{1}{d}-\beta_{b}\right)
}\right)  ^{m_{j}}%
\end{align}
Here the binomial coefficient $\binom{\left\Vert S\right\Vert }{n}$ bounds the
choice for the components $\delta_{1},...,\delta_{n}$ of $X_{b}$ to hit the
boundary layer at $i_{1},...,i_{n}$. The above inequality yields%
\begin{align}
&  \sum_{\Gamma^{p}:\text{supp}\,\Gamma^{p}=S}\varrho(\Gamma^{p}%
)\,\,q^{\underline{e}\left\Vert \Gamma^{p}\right\Vert }\leq\sum_{0\leq
k\leq\left\Vert S\right\Vert }\binom{\left\Vert S\right\Vert }{k}%
2^{2(d-1)k}2^{\left\Vert S\right\Vert -k}q^{-\frac{k}{2(d-1)}}\nonumber\\
&  \hphantom{xxxxxxx}\times\sum_{n\leq\left\Vert S\right\Vert }\binom
{\left\Vert S\right\Vert }{n}\sum_{m\geq\left\Vert S\right\Vert -k}\left(
2\nu_{d}q^{-\left(  \frac{1}{d}-\beta_{b}\right)  }\right)  ^{m}\nonumber\\
&  \hphantom{xx}\leq\sum_{0\leq k\leq\left\Vert S\right\Vert }\binom
{\left\Vert S\right\Vert }{k}2^{2(d-1)k}2^{\left\Vert S\right\Vert -k}%
q^{-\frac{k}{2(d-1)}}\nonumber\\
&  \hphantom{xxxxxxxxxxxx}\times\left(  2\nu_{d}q^{-\left(  \frac{1}{d}%
-\beta_{b}\right)  }\right)  ^{\left\Vert S\right\Vert -k}\frac{2^{\left\Vert
S\right\Vert }}{1-2\nu_{d}q^{-\left(  \frac{1}{d}-\beta_{b}\right)  }}%
\end{align}
that gives the inequality of the lemma.
\end{proof}

We now turn to the general case of contours with non empty polymers and first
give a bound on the activity $\psi\left(  \gamma\right)  $ of polymers.

\begin{lemma}
\label{AP3}\bigskip Assume that $\beta_{b}<\frac{1}{d}$, and $c\nu_{d}%
^{2}q^{-\frac{1}{d}+\beta_{b}}\leq1$ with $c=8e(e-1)(\sqrt{3}+2)e^{\frac
{2}{\sqrt{3}+1}}$ and $\nu_{d}=(2d)^{2}$, then%
\begin{equation}
\left|  \psi\left(  \gamma\right)  \right|  \leq\left(  c\nu_{d}^{2}%
q^{-\frac{1}{d}+\beta_{b}}\right)  ^{\left|  \gamma\right|  }%
\end{equation}

\end{lemma}

\begin{proof}
Let us first recall the definition (\ref{eq:3.6}): $\psi(\gamma)\equiv
\sum_{A:\text{supp}\,A=\gamma}\omega(A)$ where the weights of aggregates are
defined by (see (\ref{eq:3.3}) and (\ref{eq:3.4})): $\omega(A)=\prod
_{\gamma\in A}e^{-\Phi(\gamma)}-1$. By Theorem \ref{CE1} we know that
$\left\vert \Phi(\gamma)\right\vert \leq\left(  ec_{0}\nu_{d}q^{-\frac{1}%
{d}+\beta_{b}}\right)  ^{\left\vert \gamma\right\vert }(\leq1)$ for $q$ large
enough. Since for any $\left\vert x\right\vert \leq1$, $\left\vert
e^{-x}-1\right\vert \leq(e-1)\left\vert x\right\vert $, we have
\begin{equation}
\left\vert \Psi(\gamma)\right\vert =\left\vert e^{-\Phi(\gamma)}-1\right\vert
\leq(e-1)\left\vert \Phi(\gamma)\right\vert \leq\left(  (e-1)ec_{0}\nu
_{d}q^{-\frac{1}{d}+\beta_{b}}\right)  ^{\left\vert \gamma\right\vert }\equiv
e^{-\sigma\left\vert \gamma\right\vert }%
\end{equation}

Then,%
\begin{align}
\sum_{A:\text{supp}\,A=\gamma}\left\vert \omega(A)\right\vert  &  =\sum
_{n\geq1}\sum_{\substack{\gamma_{1},...,\gamma_{n} \\\text{supp}\,\left\{
\gamma_{1},...,\gamma_{n}\right\}  =\gamma}}\prod_{j=1}^{n}\left\vert
\Psi(\gamma_{j})\right\vert \nonumber\\
&  \leq\sum_{n\geq1}2^{\left\vert \gamma\right\vert }\sum_{\substack{\gamma
_{1}\backepsilon b_{1},...,\gamma_{n}\backepsilon b_{n} \\\text{supp}%
\,\left\{  \gamma_{1},...,\gamma_{n}\right\}  =\gamma}}\prod_{j=1}%
^{n}e^{-\sigma\left\vert \gamma_{j}\right\vert }\nonumber\\
&  \leq\sum_{n\geq1}2^{\left\vert \gamma\right\vert }\sum_{\substack{m_{1}%
,...,m_{n} \\m_{1}+...+m_{n}\geq\left\vert \gamma\right\vert }}\prod_{j=1}%
^{n}\left(  \nu_{d}e^{-\sigma}\right)  ^{m_{j}}\nonumber\\
&  \leq\sum_{n\geq1}\sum_{\substack{m_{1},...,m_{n} \\m_{1}+...+m_{n}%
\geq\left\vert \gamma\right\vert }}\prod_{j=1}^{n}\left(  2\nu_{d}e^{-\sigma
}\right)  ^{m_{j}}%
\end{align}
Here, we used as in the proof of Theorem \ref{CE1} that the number of polymers
of length $m$ containing a given bond or a given vertex is less than $\nu
_{d}{}^{m}$; the term $2^{\left\vert \gamma\right\vert }$ bounds the
combinatoric choice of the bonds or vertices $b_{j}\in\gamma_{j}$, because
$\gamma$ being connected, it contains $n-1$ such intersecting cells (see
\cite{GMM}).

We put $k=m_{1}+...+m_{n}$ and notice that there are at most $\binom{k}{n-1}$
such numbers to get
\begin{align}
\sum_{A:\text{supp}\,A=\gamma}\left\vert \omega(A)\right\vert  &  =\sum_{1\leq
n\leq k}\sum_{k\geq\left\vert \gamma\right\vert }\binom{k}{n-1}\left(
2\nu_{d}e^{-\sigma}\right)  ^{k}\nonumber\\
&  \leq\sum_{k\geq\left\vert \gamma\right\vert }\left(  4\nu_{d}e^{-\sigma
}\right)  ^{k}=\sum_{k\geq\left\vert \gamma\right\vert }\left(  \frac{c}{2}%
\nu_{d}^{2}q^{-\frac{1}{d}+\beta_{b}}\right)  ^{k}\nonumber\\
&  \leq\frac{1}{1-\frac{c}{2}\nu_{d}^{2}q^{-\frac{1}{d}+\beta_{b}}}\left(
\frac{c}{2}\nu_{d}^{2}q^{-\frac{1}{d}+\beta_{b}}\right)  ^{\left\vert
\gamma\right\vert }%
\end{align}
provided that $\frac{c}{2}\nu_{d}^{2}q^{-\frac{1}{d}+\beta_{b}}<1$. The lemma
then follows by assuming that $\frac{c}{2}\nu_{d}^{2}q^{-\frac{1}{d}+\beta
_{b}}\leq\frac{1}{2}$
\end{proof}

We finally turn to the

\textbf{Proof of Proposition \ref{P:PE1}}

Consider a contour $\Gamma=\left\{  \text{supp}\,\Gamma,X^{\Gamma},Y^{\Gamma
}\right\}  $ and as above the decomposition $X^{\Gamma}=X_{s}^{\Gamma}\cup
X_{b}^{\Gamma}$ where $X_{s}^{\Gamma}=X^{\Gamma}\cap B(\mathbb{L}_{0})$ is the
set of bonds of $X^{\Gamma}$\ on the boundary layer and $X_{b}^{\Gamma
}=X^{\Gamma}\setminus X_{s}^{\Gamma}$. Consider also the union $Z_{b}%
=X_{b}^{\Gamma}\cup Y^{\Gamma}$. Notice that the set $Z=X_{s}^{\Gamma}\cup
Z_{b}$ is a family of hydras and there are at most $3^{\left\vert
Z_{b}\right\vert }$ contours corresponding to this family: this is because a
bond in $Z_{b}$ may be occupied either by $X_{b}^{\Gamma}$ or by $Y^{\Gamma}$
or by both. Let%
\begin{equation}
\left\vert \widetilde{\varrho}(Z)\right\vert =\sum_{\Gamma:X_{b}^{\Gamma}\cup
Y^{\Gamma}=Z}\left\vert \varrho(\Gamma)\right\vert
\end{equation}
The above remark on the number of contours associated to $Z$ and Lemma
\ref{AP3}\ implies%
\begin{align}
\left\vert \varrho(\Gamma)\,\right\vert q^{\underline{e}\left\Vert
\Gamma\right\Vert }  &  \leq q^{-\frac{\left\vert I_{0}(\Gamma)\right\vert
}{2(d-1)}}\left(  3\sup\left\{  q^{-\left(  \frac{1}{d}-\beta_{b}\right)
},c\nu_{d}^{2}q^{-\frac{1}{d}+\beta_{b}}\right\}  \right)  ^{\left\vert
Z_{b}\right\vert }\nonumber\\
&  \leq q^{-\frac{\left\vert I_{0}(\Gamma)\right\vert }{2(d-1)}}\left(
3c\nu_{d}^{2}q^{-\frac{1}{d}+\beta_{b}}\right)  ^{\left\vert Z_{b}\right\vert
}%
\end{align}
The rest of the proof is then analog to that of Lemma \ref{AP2} starting from
Lemma \ref{AP1} and replacing $\nu_{d}q^{-\frac{1}{d}+\beta_{b}}$ by
$3c\nu_{d}^{3}q^{-\frac{1}{d}+\beta_{b}}$. It gives%
\begin{equation}
\sum_{\Gamma^{p}:\text{supp}\,\Gamma^{p}=S}\left\vert \varrho(\Gamma
^{p})\,\right\vert \,q^{\underline{e}\left\Vert \Gamma^{p}\right\Vert }%
\leq\left(  2^{2d-1}q^{-\frac{1}{2(d-1)}}+24c\nu_{d}^{3}q^{- \frac{1}{d}%
+\beta_{b} }\right)  ^{\left\Vert S\right\Vert }\frac{1}{1-6c\nu_{d}^{3}q^{-
\frac{1}{d}+\beta_{b} }}%
\end{equation}
provided $6c\nu_{d}^{3}q^{-\frac{1}{d}+\beta_{b}}<1$ and ends the proof of the
proposition. {\rule{0.5em}{0.5em}}

\newpage\thispagestyle{empty}

\section*{Figure captions}

\begin{enumerate}
\item The diagram of ground states.

\item A hydra, in two dimensions (a dimension not considered in this paper),
with $2$ feet (components of full lines), $5$ necks (dotted lines), and $3$
heads (components of dashed lines).
\end{enumerate}

\newpage\thispagestyle{empty}

\begin{center}
\setlength{\unitlength}{6.5mm} \begin{picture}(14,7)
\put(1,-3.5){
\begin{picture}(0,0)
\drawline(0,0)(0,7)
\drawline(0,0)(8,0)       %
\put(7.8,-0.15){$\blacktriangleright$} \put(8,-0.7){$\beta_{b}$}
\put(-0.175,6.9){$\blacktriangle$} \put(-0.7,7){$\beta_{s}$}
\put(1.9,-0.7){$\frac{1}{d}$} \put(5.9,-0.7){$1$}
\put(-1.1,2.85){$\frac{1}{d-1}$}
\drawline(0,3)(2,3) \drawline(2,2)(6,0) \drawline(2,0)(2,6.5)
\put(.5,1.5){I} \put(.5,4.5){II} \put(4.2,3.7){IV}
\put(2.5,0.5){III} \put(2.3,2.1){$S_2$} \put(2.3,3.1){$S_1$}
\end{picture}
}
\end{picture}

\end{center}

\newpage\thispagestyle{empty}

\begin{center}
\setlength{\unitlength}{8 mm} \begin{picture}(17,6)(0,0)
\drawline(0,0)(3,0) \drawline(5,0)(8,0)
\dottedline{.1}(1,0)(1,1) \dottedline{.1}(2,0)(2,1)
\dottedline{.1}(6,0)(6,1) \dottedline{.1}(8,0)(8,1)
\dottedline{.1}(10,0)(10,1)
\dashline{.1}(0,1)(1,1) \dashline{.1}(0,2)(1,2)
\dashline{.1}(1,1)(1,2) \dashline{.1}(0,1)(0,2)
\dashline{.1}(1,2)(1,3)
\dashline{.1}(2,1)(2,2) \dashline{.1}(2,1)(2,2)
\dashline{.1}(2,2)(2,3) \dashline{.1}(2,3)(4,3)
\dashline{.1}(4,2)(4,3)
\dashline{.1}(2,2)(6,2) \dashline{.1}(6,1)(6,2)
\dashline{.1}(5,2)(5,3) \dashline{.1}(5,3)(6,3)
\dashline{.1}(6,3)(6,4)
\dashline{.1}(8,1)(10,1)\dashline{.1}(9,1)(9,3)
\dashline{.1}(9,3)(11,3)\dashline{.1}(9,2)(10,2)
\dashline{.1}(10,2)(10,3)
\end{picture}

\end{center}

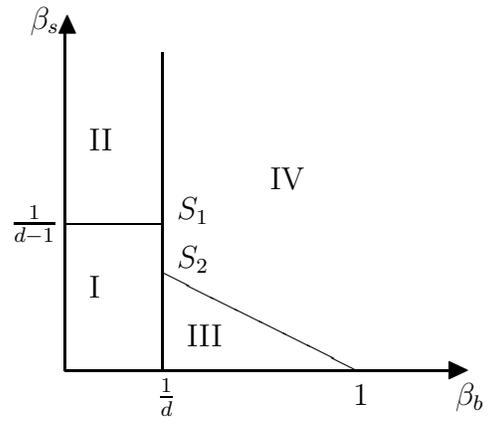
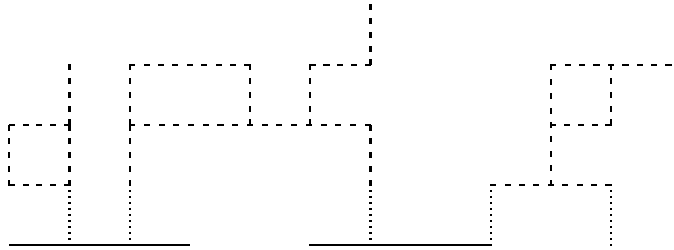
\end{document}